\documentclass[10pt]{article}
\usepackage{fancyhdr}
\usepackage{extramarks}
\usepackage{amsmath}
\usepackage{amsthm}
\usepackage{amsfonts}
\usepackage{siunitx}
\usepackage{tikz}
\usepackage[plain]{algorithm}
\usepackage{algpseudocode}
\usepackage{multirow}
\usepackage{booktabs}
\usepackage{palatino}
\usepackage{graphicx}
\usepackage{subfigure}
\usepackage[colorlinks,linkcolor=black,anchorcolor=black,citecolor=black,urlcolor=blue]{hyperref}
\usepackage{amsmath,bm}
\usepackage{booktabs}
\usepackage{mathtools}
\usepackage{amssymb}
\usepackage{caption}
\usepackage{capt-of}
\usepackage{mciteplus}
\usepackage{cite}
\usepackage{mathrsfs}
\usepackage[title,titletoc,toc]{appendix}
\usepackage{xr}
\usepackage{parskip}
\usepackage{soul}
\usepackage{textcomp}
\usepackage[colaction]{multicol}
\usepackage[switch]{lineno}
\usepackage{lipsum}
\usepackage{etoolbox}
\usepackage{longtable}
\usepackage{array}
\usepackage{tablefootnote}
\usepackage{multirow}
\newcolumntype{C}[1]{>{\centering\arraybackslash}p{#1}}
\usetikzlibrary{automata,positioning}
\usetikzlibrary{automata,positioning}

\begin{document}

\title{Characterizing  SARS-CoV-2 mutations in the United States }
 
\author{Rui Wang$^1$, Jiahui Chen$^1$,  Kaifu Gao$^1$,  Yuta Hozumi $^1$, Changchuan Yin$^2$, 
 and Guo-Wei Wei$^{1,3,4}$\footnote{
Corresponding author.		E-mail: weig@msu.edu} \\
$^1$ Department of Mathematics, \\
Michigan State University, MI 48824, USA.\\
$^2$ Department of Mathematics, Statistics, and Computer Science, \\
University of Illinois at Chicago, Chicago, IL 60607, USA\\
$^3$ Department of Electrical and Computer Engineering,\\
Michigan State University, MI 48824, USA. \\
$^4$ Department of Biochemistry and Molecular Biology,\\
Michigan State University, MI 48824, USA. \\
}
\date{\today} 

\maketitle

\begin{abstract}
The severe acute respiratory syndrome coronavirus 2 (SARS-CoV-2) has been mutating since it was first sequenced in early January 2020. The genetic variants have developed into a few distinct clusters with different properties.  Since the United States (US) has the highest number of viral infected patients globally, it is essential to understand the US SARS-CoV-2. Using genotyping,  sequence-alignment, time-evolution, $k$-means clustering, protein-folding stability, algebraic topology, and network theory, we reveal that the US SARS-CoV-2 has four substrains and five top US SARS-CoV-2 mutations were first detected in China (2 cases), Singapore (2 cases), and the United Kingdom (1 case).  The next three top US SARS-CoV-2 mutations were first detected in the US. These eight top mutations belong to two disconnected groups. The first group consisting of 5 concurrent mutations is prevailing, while the other group with three concurrent mutations gradually fades out. Our analysis suggests that female immune systems are more active than those of males in responding to SARS-CoV-2 infections. We identify that one of the top mutations, 27964C$>$T-(S24L) on ORF8, has an unusually strong gender dependence.  Based on the analysis of all mutations on the spike protein,  we further uncover that three of four US SASR-CoV-2 substrains become more infectious. Our study calls for effective viral control and containing strategies in the US. 

\end{abstract}
\pagenumbering{roman}
\begin{verbatim}
\end{verbatim}

 {\setcounter{tocdepth}{4} \tableofcontents}
  \newpage
 
\setcounter{page}{1}
\renewcommand{\thepage}{{\arabic{page}}}


\section{Introduction}
The severe acute respiratory syndrome coronavirus 2 (SARS-CoV-2), a strain of $\beta$-coronavirus that causes the respiratory illness, is responsible for the ongoing global pandemic of coronavirus disease 2019 (COVID-19). At this stage, more than 200 countries, regions, and territories have reported positive COVID-19 infections. Among them, the United States (US) has over 4 million confirmed cases and  145 thousand deceased cases as of July 22, 2020 \cite{who_2020}. The rapid spread of COVID-19 gives rise  to a question of whether SARS-CoV-2 has become more transmissible or infectious in the US. Benefiting from the vast collection of complete genome sequencing data of SARS-CoV-2 deposited in GISAID, studies on the mutation dynamics provide us a way to investigate the characteristics of US SARS-CoV-2 strains and understand their ramifications in the US population health and economy. 

SARS-CoV-2 belongs to the Coronaviridae family and the Nidovirales order, which has been shown to have a genetic proofreading mechanism in its replication achieved by an enzyme called non-structure protein 14 (NSP14) in synergy with NSP12, i.e.,  RNA-dependent RNA polymerase (RdRp) \cite{sevajol2014insights, ferron2018structural}. Therefore, SARS-CoV-2 has a higher fidelity in its transcription and replication process than that of other single-stranded RNA viruses, such as flu virus and HIV. Nonetheless, 4968 single mutations have been gradually detected in the 7823 US strains in the past few months compared with the first SARS-CoV-2 genome collected on December 24, 2019 \cite{wu2020new, wang2020decoding0}. The frequency of virus mutations is accumulated by the natural selection, cellular environment, polymerase fidelity, random genetic drift, features of recent epidemiology, host immune responses, gene editing, replication mechanism, etc \cite{sanjuan2016mechanisms, grubaugh2020making}. However, it is difficult to determine the detailed mechanism of a specific mutation. Additionally, the current mechanistic understanding of SARS-CoV-2 proteins, protein-protein interactions, and their synergy with host cell proteins,  enzymes, and signaling pathways is very limited. There is an urgent need to further explore the SARS-CoV-2 structure, function, and activity. Analyzing SARS-CoV-2 genome mutations and evolution provides an efficient means to understand viral genotyping and phenotyping. Although experimental verification is required to ultimately determine how single mutations with high frequency have changed SARS-CoV-2 properties, such as viral infectivity and virulence, the relationship between these single high-frequency mutations and SARS-CoV-2 properties can still be deciphered from genotyping, protein analysis, and transmission tracking. 

In this work, we 
 extract  7823 single nucleotide polymorphism (SNP) profiles from  genome isolates collected in the United States.  We found that the US SARS-CoV-2 has developed into four substrains. To further understand the characteristic of the US SARS-CoV-2 evolution and transmission, we investigate its most prevalent high-frequency or top mutations. Based on  genotyping results, top eight  missense mutations (i.e., 14408C$>$T-(P323L), 23403A$>$G-(D614G), 25563G$>$T-(Q57H), 1059C$>$T-(T85I), 28144T$>$C-(L84S), 17858A$>$G-(Y541C), 17747C$>$T-(P504L), and 27964C$>$T-(S24L)) are identified, in addition to three synonymous mutations  (i.e., 3037C$>$T-(F106F), 8782C$>$T-(S76S), and 18060C$>$T-(L7L))  that do not change  SARS-CoV-2 proteins.  Among them, mutations 17858A$>$G-(Y541C), 17747C$>$T-(P504L), and 27964C$>$T-(S24L) were originated and prevalent mainly in the US,  attributing to the unique characteristics of COVID-19 in the US. We focus on the top eight missense mutations in the present work. Based on co-mutation and time evolution analysis, we hypothesize that three concurrent mutations 17747C$>$T-(P504L), 17858A$>$G-(Y541C), and 28144T$>$C tend to fade out, while the other five concurrent mutations can enhance the infectivity of SARS-CoV-2. Moreover, by analyzing the gender disparity, we find that a US-unique mutation,  27964C$>$T-(S24L), shows an interesting female-dominated pattern. 

NSP2, NSP12 (RdRp), NSP13 (helicase), spike (S) protein, ORF3a, and ORF8 are six SARS-CoV-2 proteins that associate with eight different missense mutations mentioned above. It is important to analyze the impact of mutations on the structures, stabilities, and functions of these proteins.  Employing the Jaccard distance-based genotyping \cite{levandowsky1971distance,yin2020genotyping}, topological data analysis \cite{carlsson2009topology}, artificial intelligence \cite{cang2017analysis}, flexibility-rigidity index (FRI) \cite{xia2013multiscale}, and network models \cite{estrada2020topological}, we show that 23403A$>$G-(D614G) and 27964C$>$T-(S24L) strengthen the folding stability of the spike protein and ORF8 protein. Conversely, the other high-frequency missense mutations disrupt the folding stability of its corresponding proteins. Furthermore, because SARS-CoV-2 enters the host cells by the interaction of the S protein and the host angiotensin-converting enzyme 2 (ACE2) receptor \cite{hoffmann2020sars}, we quantitatively evaluate the mutation-induced protein-protein binding free energy changes ($\Delta\Delta G$) of S protein and ACE2. The results reveal that overall, the frequency of infectivity-strengthening mutations on the S protein is higher than the infectivity-weakening mutations, explaining the fast person-to-person transmission in the United States. 

In a nutshell, we analyze the characteristics of SARS-CoV-2 substrains and prevalent mutations in the United States. These mutations, together with additional mutations on the S protein, suggest that  SARS-CoV-2  has  become more infectious in the United States. 

\section{Results and Discussion}

\subsection{Genotyping analysis}
\begin{figure}[ht!]
    \centering
		\includegraphics[width=0.8\textwidth]{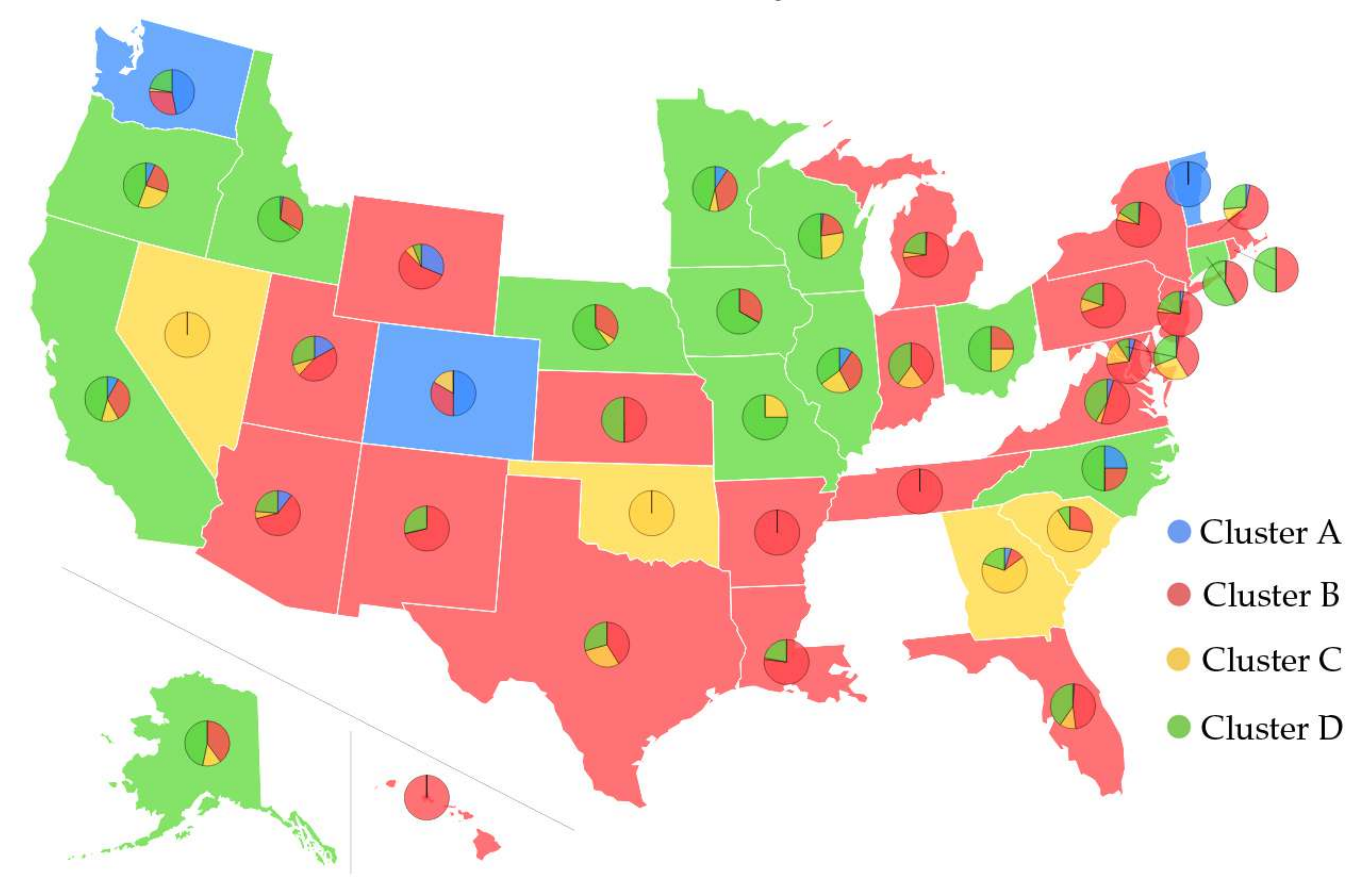}
    \caption{Pie chart plot of four clusters in the United States as of July 14, 2020. The blue, red, yellow, and green colors represent clusters A, B, C, and D, respectively. The base color of each state is decided by its dominant cluster. Some of the states do not submit the complete genome sequences to GISAID. Therefore, we will not set the base color of these states.}
    \label{fig:USMap}
\end{figure}

\subsubsection{Cluster analysis}
Complete genome sequence data can provide us with a wide variety of opportunities to decode the mutation-induced transmission and infection behavior of COVID-19. In this work, we downloaded 28726 complete SARS-CoV-2 genome sequences from GISAID (\url{https://www.gisaid.org/}) up to July 14, 2020. Based on the genotyping results, we obtain 28726 SNP profiles, which record all of the single mutations compared to the first complete genome sequence of SARS-CoV-2 in the GenBank (access number: NC\_045512.2). Among them, 7823 SNP profiles are decoded from the genome isolates submitted by the United States, and 4968 single mutations are detected. We track the geographical distributions of the 4968 single mutations in 7832 SARS-CoV-2 genome isolates with the $k$-means clustering method. Based on the mutations, the 7832 genome isolates in the United States are clustered into four distinct clusters, as shown in \autoref{fig:USMap}. The blue, red, yellow, and green represent Cluster A, B, C, and D, respectively. The base color of each state is determined by its dominated cluster. We show that most of the states are dominated by Cluster B and Cluster D. \autoref{table:US} lists the distribution of samples and the total number of single mutations in   20 US states that have submitted significantly many SARS-CoV-2 genome isolates. They are Alaska (AK), Arizona (AZ), California (CA), Connecticut (CT), Washington, D.C. (DC), Florida (FL), Idaho (ID), Illinois (IL), Louisiana (LA), Maryland (MD), Massachusetts (MA), Michigan (MI), Minnesota (MN), New Mexico (NM), New York (NY), Oregon (OR), Utah (UT), Virginia (VA), Washington (WA), and Wisconsin (WI). In Cluster A, B, C, and D, the co-mutations with the highest number of descendants are [8782C$>$T, 18060C$>$T, 28144T$>$C], [241C$>$T, 3037C$>$T, 14408C$>$T, 23403A$>$G], [11083G$>$T], and [3037C$>$T, 14408C$>$T], respectively. It is noted that all of the 20 states have the mutations from Clusters B and D. Alaska (AK) and Louisiana (LA) do not have mutations from  Cluster A, and New Mexico (NM) does not have mutations in Clusters A and C. The complete list of the table can be found in the supporting materials. More analysis related to the infectivity of SARS-CoV-2 based on our four distinct clusters is given in  Section \ref{sec:Infectivity}.

\begin{table}[ht]
    \centering
    \setlength\tabcolsep{11pt}
    \captionsetup{margin=0.9cm}
    \caption{The cluster distributions of samples ($N_{\rm NS}$) and total mutation counts ($N_ {\rm TF}$) in 20 states of the U.S. The 20 states are Alaska (AK), Arizona (AZ), California (CA), Connecticut (CT), Washington, D.C. (DC), Florida (FL), Idaho (ID), Illinois (IL), Louisiana (LA), Maryland (MD), Massachusetts (MA), Michigan (MI), Minnesota (MN), New Mexico (NM), New York (NY), Oregon (OR), Utah (UT), Virginia (VA), Washington (WA), and Wisconsin (WI).}
    \begin{tabular}{l|cc|cc|cc|cc}
    \hline
     &   \multicolumn{2}{c|}{Cluster A}    & \multicolumn{2}{c|}{Cluster B}    & \multicolumn{2}{c|}{Cluster C}    & \multicolumn{2}{c}{Cluster D}      \\ \hline
         State                &   $N_{\rm NS}$ &$N_{\rm TF}$  & $N_{\rm NS}$ &$N_{\rm TF}$  & $N_{\rm NS}$ &$N_{\rm TF}$  & $N_{\rm NS}$ &$N_{\rm TF}$ \\  \hline
        AK&0&0&12&99&4&27&14&105\\
        AZ&8&51&46&364&4&44&18&156\\
        CA&99&752&414&3308&152&830&561&5999\\
        CT&1&11&44&365&1&12&62&701\\
        DC&1&6&15&108&4&51&2&34\\
        FL&2&12&85&672&21&275&73&810\\
        ID&1&10&16&147&1&9&33&460\\
        IL&8&57&28&202&19&184&30&265\\
        LA&0&0&264&2302&3&80&77&961\\
        MD&1&6&23&178&16&221&18&257\\
        MA&1&8&21&138&3&21&9&64\\
        MI&2&13&192&1486&11&111&61&572\\
        MN&43&311&177&1398&31&236&212&2537\\
        NM&0&0&15&98&0&0&6&52\\
        NY&17&108&977&7338&70&560&204&1918\\
        OR&13&84&46&395&50&269&87&762\\
        UT&12&76&32&243&6&50&21&161\\
        VA&15&100&163&1291&13&108&136&1430\\
        WA&982&7274&598&4799&49&558&465&5316\\
        WI&11&74&137&1019&178&973&334&3539\\
    \hline
\end{tabular}
\label{table:US}
\end{table}

\subsubsection{Top mutations in the United States}

\begin{table}[ht]   
    \caption{Top 8 missense mutations that are prevalent in the United States. The ranking of these 8 mutations in the US and world are included in the table. NC$_{\text{U.S.}}$ and NC$_{\text{World}}$ represent for the total number of sequences with a specific mutation in the United States and in the world, respectively. The last column records the date that these eight missense mutations were detected for the first time in the world and in the United States. The second-last column records their corresponding countries, i.e. the country lists at the top shows where the mutations first detected, and the country lists at the bottom will always be the United States. Here, UK, US, CN, SG represent the United Kingdom, the United States, China, and Singapore, respectively. We use ISO 8601 format YYYY-MM-DD as the date format.}
    \label{tab:Top 8}
    \centering
    \setlength\tabcolsep{2pt}
	\captionsetup{margin=0.1cm}
    \begin{tabular}{c|c|c|c|c|c|cccccsc}
        \hline
        Rank: US/World  & Mutation & Protein & NC$_{\text{U.S.}}$ & NC$_{\text{World}}$ & Country & Date\\ \hline
        \multirow{2}{*}{Top 1/Top 2}  & \multirow{2}{*}{14408C$>$T-(P323L)} & \multirow{2}{*}{NSP12(RdRp)} & \multirow{2}{*}{5918} & \multirow{2}{*}{22081} & UK & 2020-02-03 \\ & & & & & US & 2020-02-28  \\\hline
        
        \multirow{2}{*}{Top 2/Top 1}  &\multirow{2}{*}{23403A$>$G-(D614G)} & \multirow{2}{*}{Spike} & \multirow{2}{*}{5912} & \multirow{2}{*}{22162} & CN & 2020-01-24 \\ & & & & & US & 2020-02-28  \\\hline
        
        \multirow{2}{*}{Top 3/Top 3} &\multirow{2}{*}{25563G$>$T-(Q57H)} & \multirow{2}{*}{ORF3a} & \multirow{2}{*}{4827} & \multirow{2}{*}{8326} & SG & 2020-02-16 \\ & & & & & US & 2020-02-29  \\\hline
        
        \multirow{2}{*}{Top 4/Top 7}  &\multirow{2}{*}{1059C$>$T-(T85I)} & \multirow{2}{*}{NSP2} & \multirow{2}{*}{4237} & \multirow{2}{*}{6595} & SG & 2020-02-16 \\ & & & & & US & 2020-02-29  \\\hline
        
        \multirow{2}{*}{Top 5/Top 8} &\multirow{2}{*}{28144T$>$C-(L84S)} & \multirow{2}{*}{ORF8} & \multirow{2}{*}{1434} & \multirow{2}{*}{2487} & CN & 2020-01-05 \\ & & & & & US & 2020-01-19  \\\hline
        
        \multirow{2}{*}{Top 6/Top 10} &\multirow{2}{*}{17858A$>$G-(Y541C)} & \multirow{2}{*}{NSP13(Helicase)} & \multirow{2}{*}{1245} & \multirow{2}{*}{1428} & US & 2020-02-20 \\ & & & & & US & 2020-02-20  \\\hline
        
        \multirow{2}{*}{Top 7/Top 11} &\multirow{2}{*}{17747C$>$T-(P504L)} & \multirow{2}{*}{NSP13(Helicase)} & \multirow{2}{*}{1224} & \multirow{2}{*}{1396} & US & 2020-02-20 \\ & & & & & US & 2020-02-20  \\\hline
        
        \multirow{2}{*}{Top 8/Top 14}  &\multirow{2}{*}{27964C$>$T-(S24L)} & \multirow{2}{*}{ORF8} & \multirow{2}{*}{705} & \multirow{2}{*}{749} & US & 2020-02-20 \\ & & & & & US & 2020-02-20  \\\hline
    \end{tabular}
\end{table}

To investigate the implications of mutations on the transmission, infection, and virulence of SARS-CoV-2 in the United States, we focus on the high-frequency or top mutations that represent the most common characteristics of SARS-CoV-2 in the United States. A total of 11 mutations in the United States has a frequency greater than 700. Among them, 3 mutations are synonymous ones (i.e., 3037C$>$T-(F106F), 8782C$>$T-(S76S), and 18060C$>$T-(L7L)) and 8 mutations are the missense mutations (i.e., 14408C$>$T-(P323L), 23403A$>$G-(D614G), 25563G$>$T-(Q57H), 1059C$>$T-(T85I), 28144T$>$C-(L84S), 17858A$>$G-(Y541C), 17747C$>$T-(P504L), and 27964C$>$T-(S24L)). Since synonymous mutations do not change SARS-CoV-2 proteins, we only focus on the other eight missense mutations. \autoref{tab:Top 8} lists the frequencies of the top 8 missense mutations in the United States. The dates that these mutations were first detected in the world and in the United States are also included in the \autoref{tab:Top 8}. The first missense mutation, 23403A$>$G-(D614G), occurred in China on January 24, 2020. The missense mutation with the highest frequency,  14408C$>$T-(P323L), occurred in the United Kingdom on February 3, 2020. Both mutations were first detected in the US on February 28, 2020. The first top mutation recorded in the US was 28144T$>$C-(L84S), on January 19. This mutation was originally detected in China on January 5, 2020. 

Note that three of the top 8 missense mutations, i.e., 17858A$>$G-(Y541C), 17747C$>$T-(P504L), and 27964C$>$T-(S24L), first appeared in the United States. In fact, over 87\% of these mutations are kept in the United States. 

Note that the top 5 mutations in the United States are also the top 5 mutations in the world. However, the next 3 mutations in the US is not ranked within the top 8 globally.

\subsubsection{Co-mutation analysis}
\begin{table}[ht!]
    \centering
    \setlength\tabcolsep{1pt}
    \captionsetup{margin=0.9cm}
    \caption{The statistical values of pairwise co-mutations from the top 8 high-frequency mutations. The values in the diagonal reveal the total number of a specific single mutation in the United States, the values in the upper triangular represent the total number of the co-mutations, and the values in the lower triangular present the ratios of pairwise co-mutations over single mutations. }
    \begin{tabular}{l|c|c|c|c|c|c|c|c}
    \hline
    &14408\text{C}$>$\text{T}&23403\text{A}$>$\text{G}&1059\text{C}$>$\text{T}&25563\text{G}$>$\text{T}&27964\text{C}$>$\text{T}&17747\text{C}$>$\text{T}&17858\text{A}$>$\text{G}&28144\text{T}$>$\text{C} \\  \hline
    14408\text{C}$>$\text{T}&5918&5904&4235&4818&705&0&0&4\\
    23403\text{A}$>$\text{G}&1.00/1.00&5912&4230&4818&705&0&0&3\\
    1059\text{C}$>$\text{T}&1.00/0.72&1.00/0.72&4237&4235&703&0&0&4\\
    25563\text{G}$>$\text{T}&1.00/0.82&1.00/0.82&1.00/0.88&4827&703&1&1&4\\
    27964\text{C}$>$\text{T}&1.00/0.12&1.00/0.12&1.00/0.17&1.00/0.15&705&0&0&0\\
    17747\text{C}$>$\text{T}&0/0&0/0&0/0&0/0&0/0&1224&1224&1222\\
    17858\text{A}$>$\text{G}&0/0&0/0&0/0&0/0&0/0&1.00/1.00&1245&1243\\
    28144\text{T}$>$\text{C}&0/0&0/0&0/0&0/0&0/0&0.85/1.00&0.87/1.00&1434\\
    \hline
\end{tabular}
\label{table:Matrix}
\end{table}
 The statistical values of pairwise co-mutations from the top 8 high-frequency mutations in \autoref{table:Matrix}. The upper triangular reveals the total number of co-mutations for each pair of mutations, the diagonal presents the frequency of every single mutation,  and the lower triangular shows the ratios of pairwise co-mutations over single mutations. It is easy to see that the top 8 mutations can be grouped into two essentially disconnected groups. The first group involves 5 mutations 1059C$>$T-(T85I), 14408C$>$T-(P323L), 23403A$>$G-(D614G), 25563G$>$T-(Q57H), and  27964C$>$T-(S24L) that are strongly correlated, though have a wide range of frequencies.  The other three mutations, 17747C$>$T-(P504L), 17858A$>$G-(Y541C), and 28144T$>$C-(L84S), occur mostly together and have similar numbers of frequencies.

\subsubsection{Evolutionary analysis}

\begin{figure}[ht!]
    \centering
	 \includegraphics[width=1\textwidth]{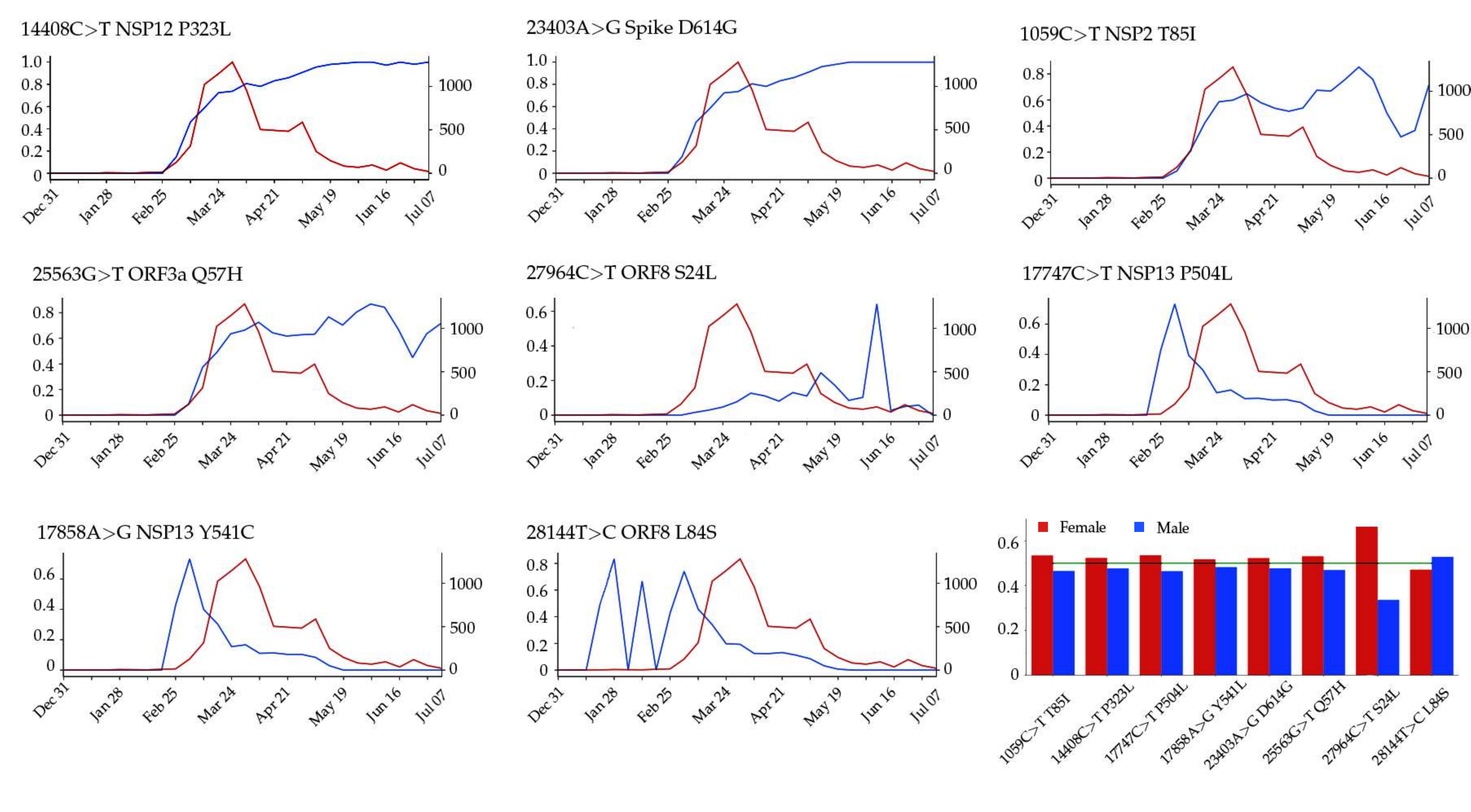}
    \caption{The blue line plots illustrate the evolution of the top 8 missense mutation ratios computed as the frequency of genome sequences having mutations over the counts of genome sequences at each 10-days period. The red lines represent the evolution of the total counts of genome sequences. Bar plot of the gender distributions of the ratio of the number of samples having top 8 missense mutations over the total number of samples having age and/or gender labels. Red bars represent the female ratio and the blue bars represent the male ratio in the United States. }
    \label{fig:Barplot}
\end{figure}

Figure \ref{fig:Barplot} plots time evolution trajectories of top 8 missense mutations. The red curves are the total numbers of genome samples over time, which become very insufficient after middle May 2020. 
First, as shown in  \autoref{table:Matrix},  mutations 14408C$>$T-(P323L) and 23403A$>$G-(D614G)  appear concurrently and thus have an identical trajectory as shown in Figure \ref{fig:Barplot}. Note that this pair of mutations exist in essentially all of the US infections.  Additionally, mutation 1059C$>$T-(T85I) always occurs together with mutation 25563G$>$T-(Q57H). Therefore, its time evolution trajectory is extremely similar to that of 25563G$>$T-(Q57H). Both mutations were first detected in Singapore on February 16, 2020. This pair of mutations occur in about 70\% of US COVID-19 cases. The third pair of mutations, 17747C$>$T-(P504L) and 17858A$>$G-(Y541C), first detected and occurred mostly in the US,  have an identical evolution trajectory. Suggested by genome samples,  this pair of US-based mutations on the helicase protein appears to have essentially died out in the US.  Unfortunately, because there are very insufficient sequencing in the US now as shown by the red curve in  Figure \ref{fig:Barplot}, one can not rule out the existence of these mutations. Mutation  28144T$>$C-(L84S), the first known mutation globally, has had a very unsteady trajectory. However, its trajectory became identical to those of its co-mutations 17747C$>$T-(P504L) and 17858A$>$G-(Y541C) after February 20, 2020.  Finally, mutation  27964C$>$T-(S24L) has an usually behavior. Its peak ratios occurred in early June when there were insufficient sequence samples in the US.

\subsubsection{Gender analysis}

The last chart in Figure \ref{fig:Barplot} displays the gender disparity of the top 8 mutations in the US. 
The overall pattern may correlate with the disparity in male and female immune response and gene editing strengths. In the US,  the total number of mutations in all 1044 female genome samples is 8438 compared with 7779 mutations recorded in 982 male samples. After the total number correction, female genome isolates still have about 2\% more mutations than those of male genome isolates. There is an apparent gender difference in mutation 27964C$>$T-(S24L) on the ORF8 protein. Among 705 samples having this mutation, 204 isolates have gender labels. The ratio of female samples (167) over male samples (67) is about 2 to 1. We currently do not have a good explanation for such a large difference in the ratio. 

\subsection{Protein-specific analysis}

In this section, we discuss the properties of the top 8 missense mutations associated with  6 proteins (i.e., NSP2, NSP12, NSP13, spike protein, ORF3a, and ORF8). Based on the analysis presented above, we reveal the potential influence of these high-frequency mutations on the infective capacity of SARS-CoV-2 in the United States using protein folding stability analysis. Furthermore, to understand the impact on the protein's structures induced by  mutations, we employ artificial intelligence \cite{cang2017analysis}, flexibility-rigidity index (FRI) \cite{xia2013multiscale}, and subgraph centrality  models \cite{estrada2020topological}. Results of our theoretical analysis are summarized in  \autoref{tab:folding stability changes}, which lists the folding stability changes, the FRI rigidity changes,  and the average subgraph centrality changes following the mutations. Here, the folding stability change following mutation $\Delta\Delta G = \Delta G_w - \Delta G_m$ measures the difference between the folding free energies of the wild type $\Delta G_w$ and the mutant type  $\Delta G_w$. More specifically, a positive folding stability change $\Delta\Delta G$ indicates that the mutation will stabilize the structure of the protein and vice versa. The molecular FRI rigidity  $R_\eta$ measures the topological connectivity and the geometric compactness of the network consisting of C$_\alpha$ at each residue and the heavy atoms involved in the mutant. In this work, $\eta$ determines the range of pairwise interactions and is set to  8\AA. Furthermore, the average subgraph centrality $\langle C_s \rangle$ is employed to describe the connection between any pair of atoms within a cutoff distance of 8.0 \AA. The decrease of the average participating rate of each edge in the network will cause the increment of the average subgraph centrality $\langle C_s \rangle$. The increment of the $\langle C_s \rangle$ induced by a mutation is due to the decrease of the number of neighbor edges participating rate of each edge \cite{estrada2020topological}. In addition, we also list the statistical values of pairwise co-mutations for the top 8 high-frequency mutations in \autoref{table:Matrix}. 
 
\begin{table}[ht]   
\caption{The protein folding stability changes of 8 missense mutations. The folding stability change $\Delta\Delta G = \Delta G_{w} - \Delta G_{m}$, where $\Delta G_{w}$ and $\Delta G_{m}$ are the folding free energies  of the wild type and  the mutant type, respectively.  $R_8^{w}$ and $R_8^{m}$ are  FRI rigidities for the wild type and mutant type of the protein with $\eta=8$\AA. Here, $\langle C_s^w \rangle$  and $\langle C_s^m \rangle$ are the average subgraph centralities of the wild type and the mutant type, respectively. $\Delta \bar{R}_{8}$ and $\Delta \langle \bar{C}_s \rangle $ are the molecular FRI rigidity changes and the average subgraph centrality change.}
\label{tab:folding stability changes}
\centering
\setlength\tabcolsep{2pt}
\captionsetup{margin=0.1cm}
\begin{tabular}{lllccccccc}
\hline
Rank &Mutation     &Protein        &$\Delta\Delta G$(kcal/mol)&$R_{8}^w$&$R_{8}^m$& $\Delta \bar{R}_{8}$\% & $\langle C_s^w \rangle$&$\langle C_s^m \rangle$ & $\Delta \langle \bar{C}_s \rangle $\%\\ \hline
Top 1&14408C$>$T-(P323L)&NSP12(RdRp)    &-0.11 & 9.77 & 9.87 &-1.0 &1105  &  1959 &-77\\
Top 2&23403A$>$G-(D614G)&S protein          & 0.34 &10.27 &10.10 &1.7  &2376  &  1386 &42\\ 
Top 3&25563G$>$T-(Q57H) &ORF3a              &-0.24 &11.33 &11.66 &-1.5 &25061 & 58592 &-134\\ 
Top 4&1059C$>$T-(T85I)  &NSP2                   &-0.05 &12.37 &12.51 &-1.1 &89764 &166399 &-85\\
Top 5&28144T$>$C-(L84S) &ORF8                &-0.99 &12.28 &12.05 &1.9 &12810 &  6504 &49\\ 
Top 6&17858A$>$G-(Y541C)&NSP13(Helicase)&-0.81 &11.52 &10.40 &9.7  &506640&  7271 &99\\
Top 7&17747C$>$T-(P504L)&NSP13(Helicase)&-0.59 & 7.52 & 7.54 &0.3  &  4668&  6094 &-31\\
Top 8&27964C$>$T-(S24L) &ORF8               & 0.20 &11.72 &11.66 &0.5  & 11685& 29777 &-155\\
\hline
\end{tabular}
\end{table}

\subsubsection{Mutation on the NSP12 protein}
\begin{figure}[ht!]
    \centering
    \includegraphics[width=0.85\textwidth]{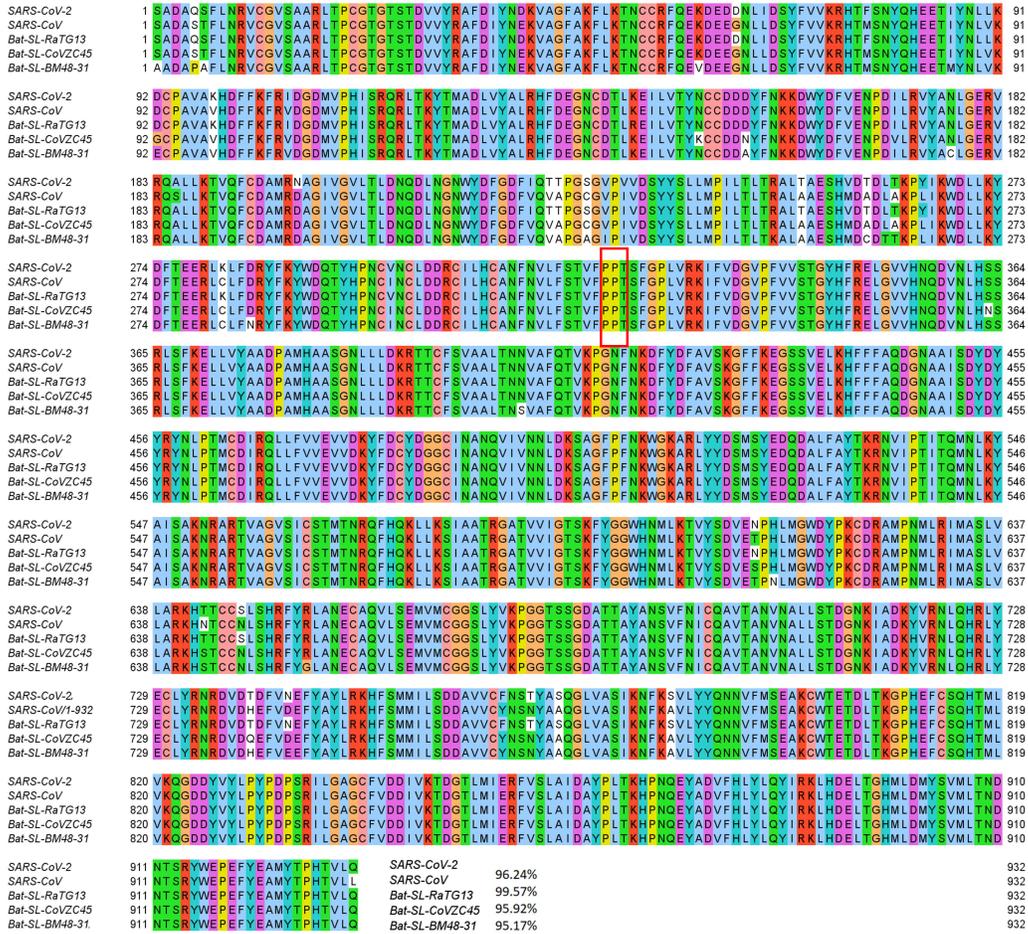}
    \caption{Sequence alignments for the NSP12 of SARS-CoV-2, SARS-CoV, bat coronavirus RaTG13, bat coronavirus CoVZC45, bat coronavirus BM48-31. Detailed numbering is given according to SARS-CoV-2. One high-frequency mutation 14408C$>$T-(P323L) is detected on the NSP12 protein. Here, the red rectangle marks the P323L mutations with its neighborhoods.}
    \label{fig:NSP12_Align}
\end{figure}

\begin{figure}[ht!]
    \centering
    \includegraphics[width=0.85\textwidth]{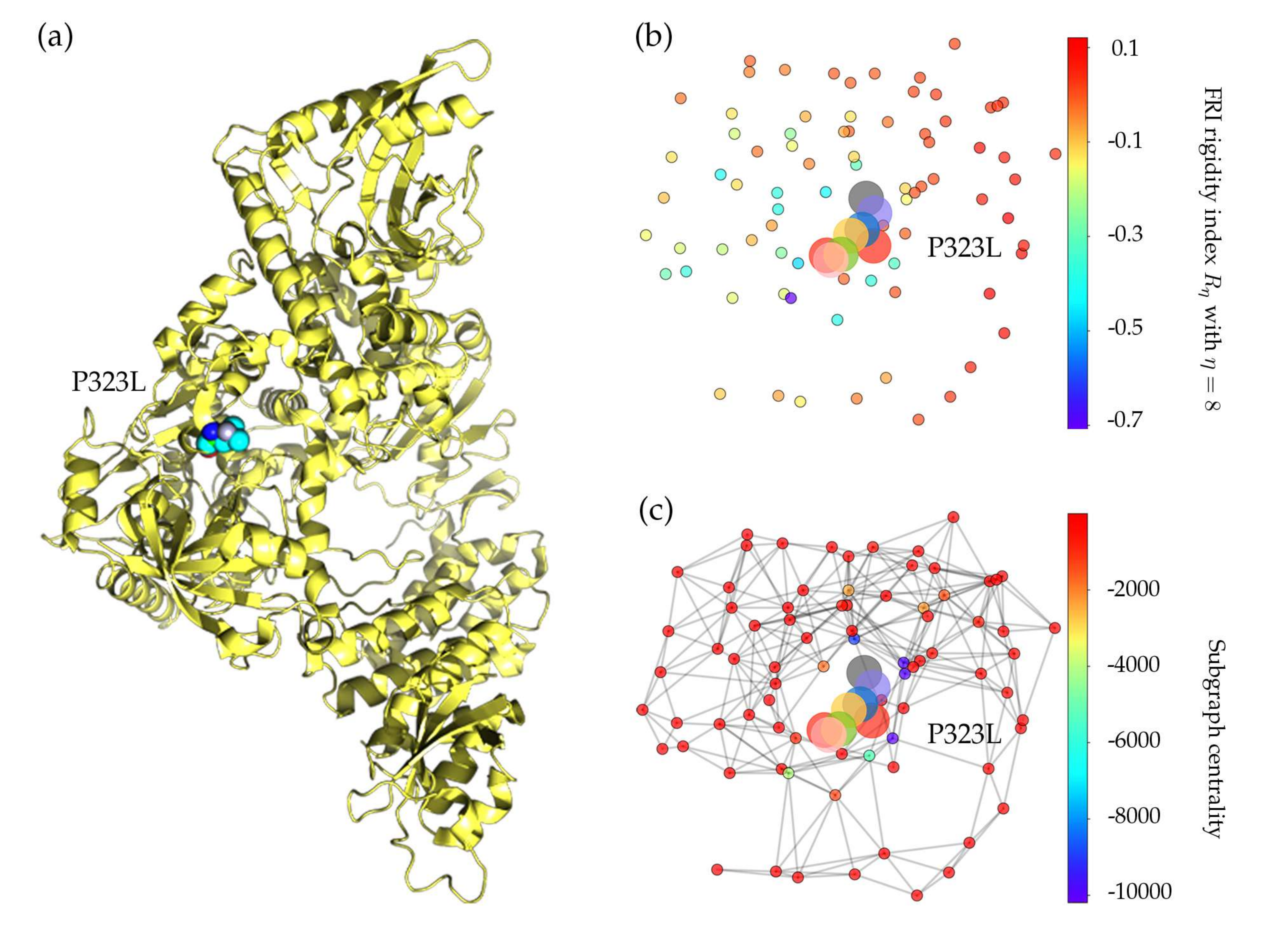}
    \caption{(a) The 3D structure of SARS-CoV-2 NSP12 protein. The mutated residue is marked with color balls. (b) The differences of FRI rigidity index between the network with wild type and the network with mutant type. (c) The difference of the subgraph centrality between the network with wild type and the network with mutant type.}
    \label{fig:NSP12_Pro}
\end{figure}

Mutation 14408C$>$T-(P323L) on the NSP12 (also called RNA-dependent RNA polymerase, abbreviation RdRp) was first found in the United Kingdom on February 03, 2020. The United States found its first case related to 14408C$>$T-(P323L) on February 28, 2020. Since then, this mutation has become one of the single dominant mutations in the United States. Among 7823 complete genome sequences, 5918 are connected to P323L. \autoref{fig:NSP12_Align} shows the sequence alignment for the NSP12 of SARS-CoV-2, SARS-CoV \cite{lee2003major},  bat coronavirus RaTG13 \cite{zhou2020pneumonia}, bat coronavirus CoVZC45\cite{hu2018genomic}, and bat coronavirus BM48-31 \cite{drexler2010genomic}. The red rectangle marks the mutant residue with its neighborhoods. One can see that SARS-CoV-2 NSP12 is highly conservative among the other four species. Although P323L mutates the residue of proline (P) to leucine (L), these two residues are both non-polar and aliphatic, indicating P323L may not affect the functionality of NSP12 too much. \autoref{fig:NSP12_Pro} (a) shows the three-dimensional (3D) structure of SARS-CoV-2 NSP12, which is created by PyMol \cite{delano2002pymol}.

 NSP12 is of paramount importance to the SARS-CoV-2 replication and transcription machinery. It is one of the primary targets for antiviral drugs. Researchers have engaged in developing new antiviral therapeutics that target NSP12 for combating COVID-19 in the past few months. NSP12 and its cofactors NSP7 and NSP8 can form a hollow cylinder-like supercomplex that interacts with NSP14, an exonuclease that has the proofreading capability \cite{pachetti2020emerging}. The structure of NSP12 contains a nidovirus-specific N-terminal extension domain (residues D60 to R249) and a right-hand RdRp domain (S367 to F920), which are connected by an interface domain (residues A250 to R365) \cite{gao2020structure}. Mutation P323L  locates on this interface domain, which,  however,  is still poorly characterized. 

The increasing ratio of P323L in \autoref{fig:Barplot} indicates that this type of mutations may favor SARS-CoV-2 and enhance the transmission capacity of SARS-CoV-2. However, the negative folding stability changes in Table \ref{tab:folding stability changes} suggest that P323L destabilizes the NSP12, which may make SARS-CoV-2 less contagious.  \autoref{fig:NSP12_Pro} (b) and (c) show the differences of FRI rigidity index and subgraph centrality between the network with wild type and the network with mutant type. The atoms on the mutant residue is mark with big color balls. We deduce that the slight increase in the rigidity means the mutation makes the protein less flexible or less cooperative in synergistic interactions. 

Based on the statistical values of co-mutations in \autoref{table:Matrix}, we can see that 14408C$>$T-(P323L) always shows up with 1059C$>$T-(T85I), 23403A$>$G-(D614G), 25563G$>$T-(Q57H), and 27964C$>$T-(S24L) simultaneously. Therefore, we can deduce that the increasing tendency of P323L ratios per 10-days is due to its co-mutation with other infectivity-strengthening mutations, such as 23403A$>$G-(D614G).

\subsubsection{Mutation on the Spike protein}
 Mutation 23403A$>$G-(D614G) located on the spike protein has the second-highest frequency in the United States, which has been recently considered as the key mutation that makes SARS-CoV-2 more infectious worldwide \cite{korber2020tracking}. From \autoref{tab:Top 8}, one can see that mutation D614G was initially detected in China on January 24, 2020. However, D614G was not widely spread in China, which is probably due to the strict social distancing rules and the Wuhan lockdown implemented by the Chinese government. In the middle of February, the hot spot with the most COVID-19 cases shifted from China to Europe and the D614G variant rapidly pervades in Europe \cite{grubaugh2020making}. The first case with the D614G mutation in the United States was reported on February 28, 2020. Since then, the D614G mutation has become a majority variant, and 68.7\% of patients carry D614G in the United States as of July 14, 2020.

\begin{figure}[ht!]
    \centering
		    \includegraphics[width=0.85\textwidth]{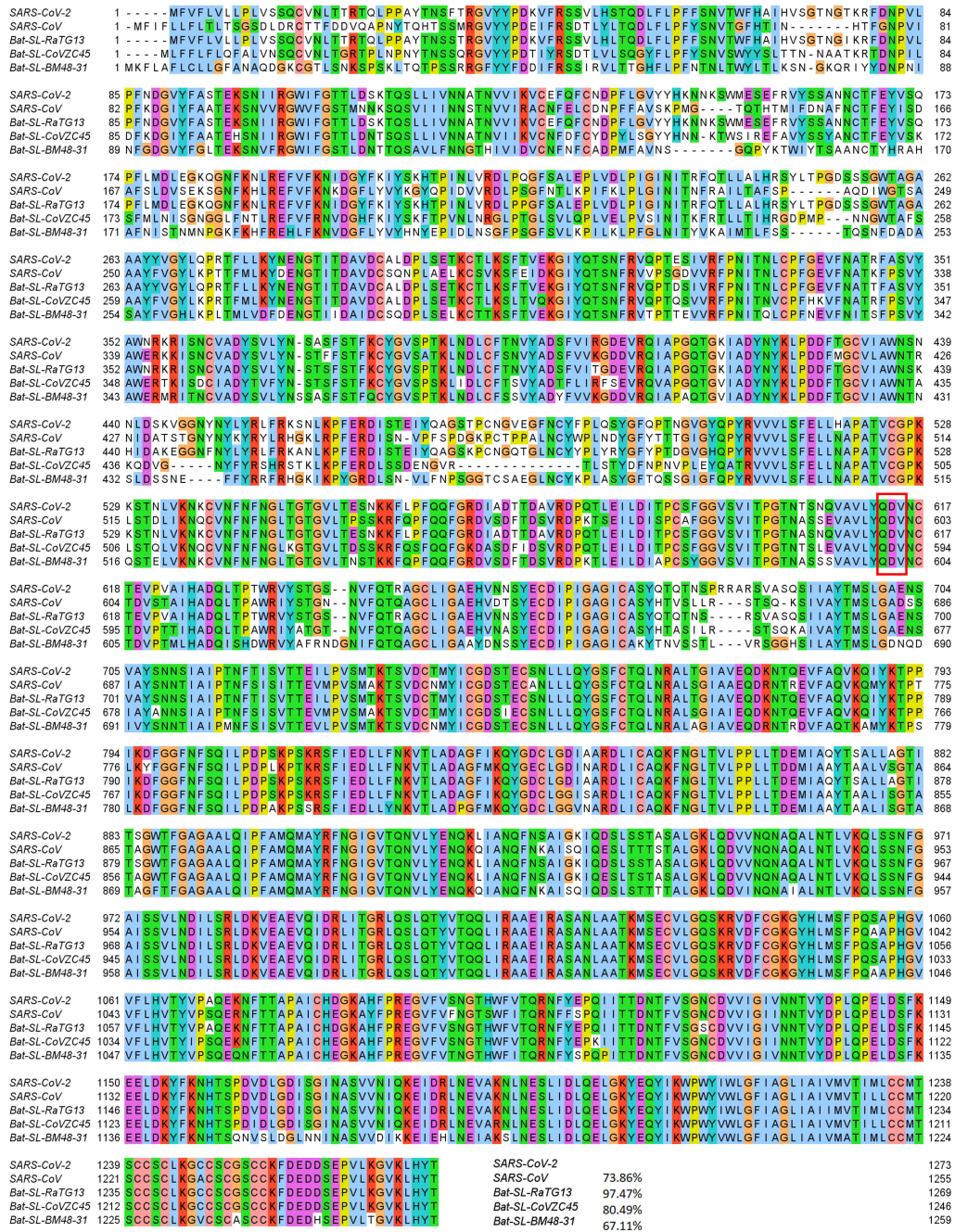}
    \caption{Sequence alignments for the S proteins of SARS-CoV-2, SARS-CoV, bat coronavirus RaTG13, bat coronavirus CoVZC45, bat
coronavirus BM48-31. Detailed numbering is given according to SARS-CoV-2 S protein. One high-frequency mutation 23403A$>$G-(D614G) is
detected on the S protein. Here, the red rectangle marks the D614G mutations with its neighborhoods}
    \label{fig:Spike_Align}
\end{figure}

\begin{figure}[ht!]
    \centering
    \includegraphics[width=0.8\textwidth]{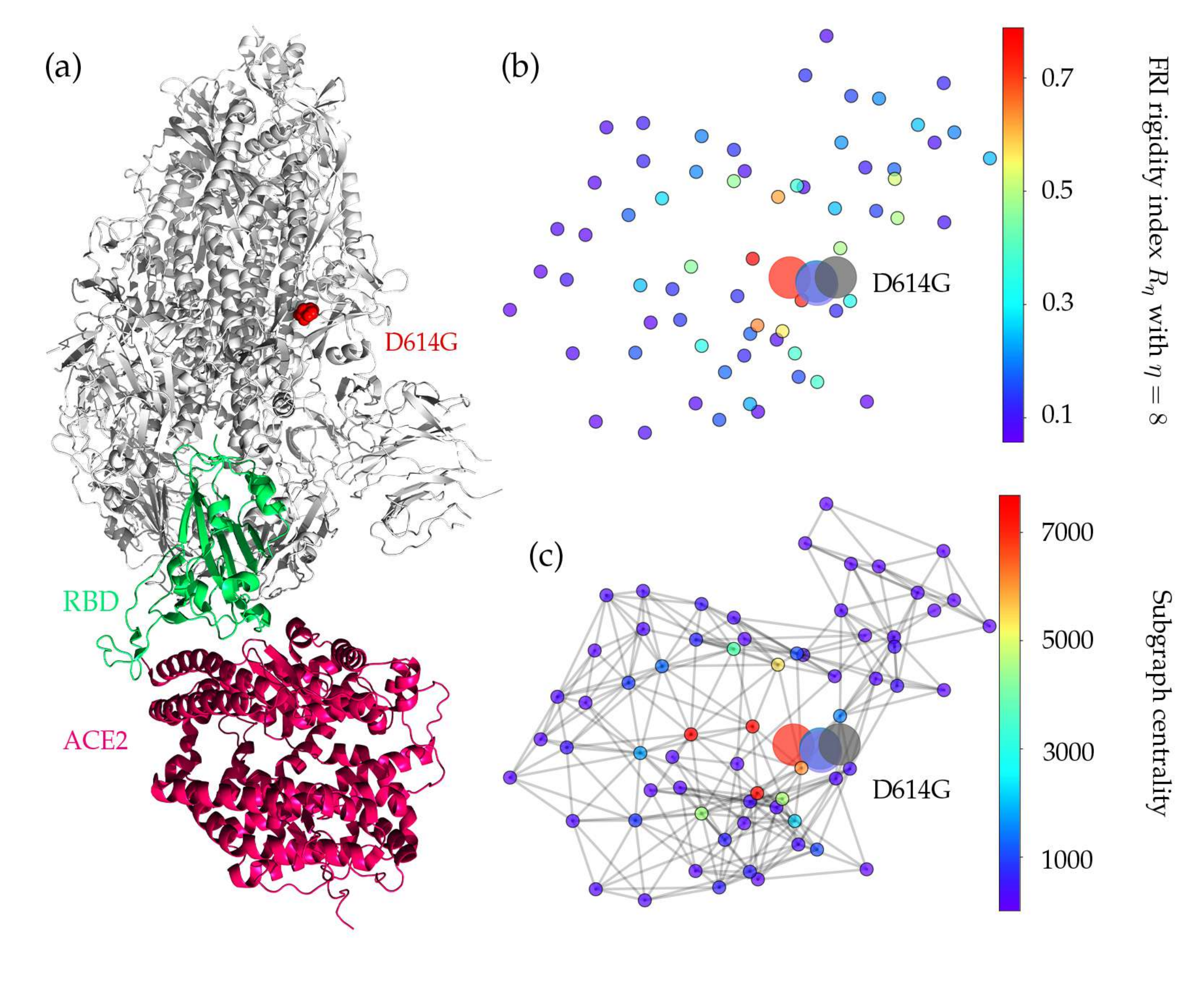}
    \caption{ Illustration of S-protein and ACE2 interaction. The RBD is displayed in green, the ACE2 is given in red, and mutation D614G is highlighted in red. (b) The difference of FRI rigidity index between the network with wild type and the network with mutant type. (c) The difference of the subgraph centrality between the network with wild type and the network with mutant type.}
    \label{fig:Spike_ACE2}
\end{figure}

The SARS-CoV-2 spike protein is a multi-functional molecule that interacts with the ACE2, which mediates the virus entering into the host cells \cite{lan2020structure}. \autoref{fig:Spike_ACE2} (a) depicts the 3D structure of the SARS-CoV-2 spike protein that interacts with the host ACE2. The D614G mutation is one of the most popular mutations of SARS-CoV-2, which changes the amino acid aspartate (D) with the polar negative charged side changes to the amino acid glycine (G) with a non-polar side chain. \autoref{fig:Spike_Align} depicts the sequence alignment for the S protein of SARS-CoV-2, SARS-CoV, bat coronavirus RaTG13, bat coronavirus CoVZC45, and bat coronavirus BM48-31. We use the red rectangle to mark the position of D614G and its neighborhoods. The S protein of bat coronavirus RaTG13 has the highest similarity of 97.47\% with the S protein of SARS-CoV-2. One can see that amino acids near position 614 are very conservative, indicating that D614G mutation will play an important role in the functions of the S protein of SARS-CoV-2. Moreover, the D614G mutation ratio in \autoref{fig:Barplot} keeps climbing, and the ratio is approaching the unity after June 16, 2020, which also proves that SARS-CoV-2  becomes more contagious as time goes on. 

 Table \ref{tab:folding stability changes} shows a positive folding free energy change, indicating the stabilization effect of the mutation.    \autoref{fig:Spike_ACE2} (b) and (c) illustrate the difference of FRI rigidity index and the subgraph centrality between the network with wild type and the network with mutant type.   The FRI rigidity decreases  following the mutation, endowing the S protein higher flexibility to interact with ACE2. The same is confirmed by average subgraph centrality.  

\subsubsection{Mutation on the ORF3a protein}
\begin{figure}[ht!]
    \centering
    \includegraphics[width=1\textwidth]{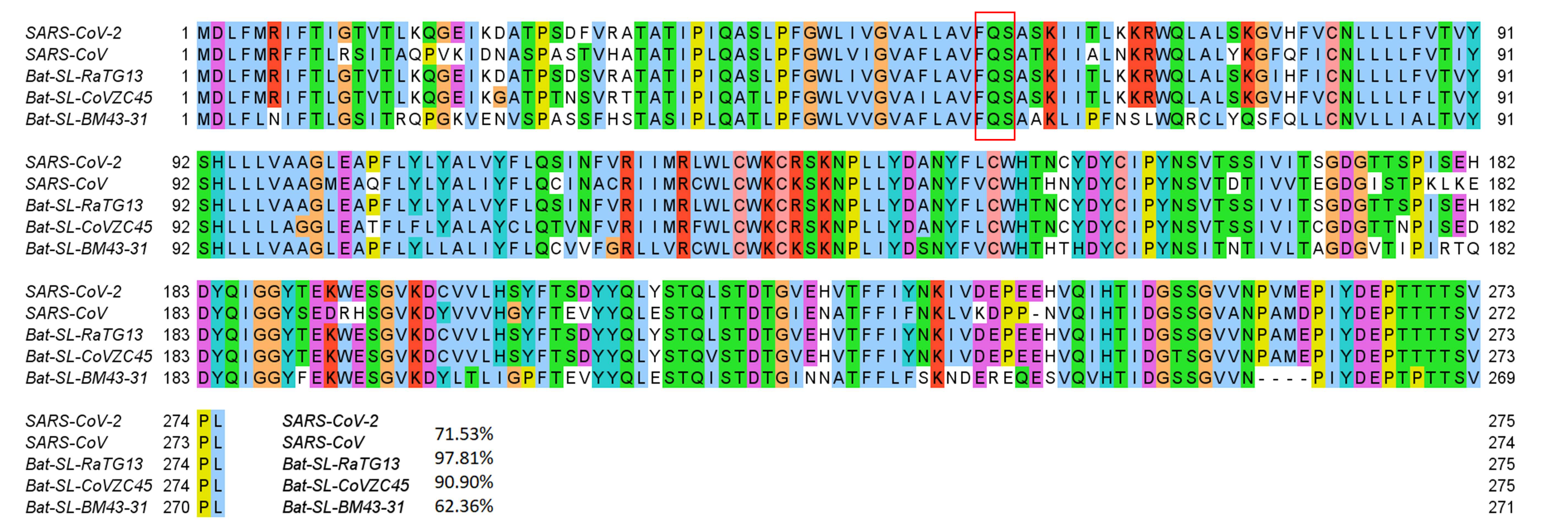}
    \caption{Sequence alignments for the ORF3a protein of SARS-CoV-2, SARS-CoV, bat coronavirus RaTG13, bat coronavirus CoVZC45, bat coronavirus BM48-31. Detailed numbering is given according to SARS-CoV-2. One high-frequency mutation 25563G$>$T-(Q57H) locates on the ORF3a protein. Here, the red rectangle marks the Q57H position with its neighborhoods.}
    \label{fig:ORF3a_Align}
\end{figure}

\begin{figure}[ht!]
    \centering
    \includegraphics[width=1\textwidth]{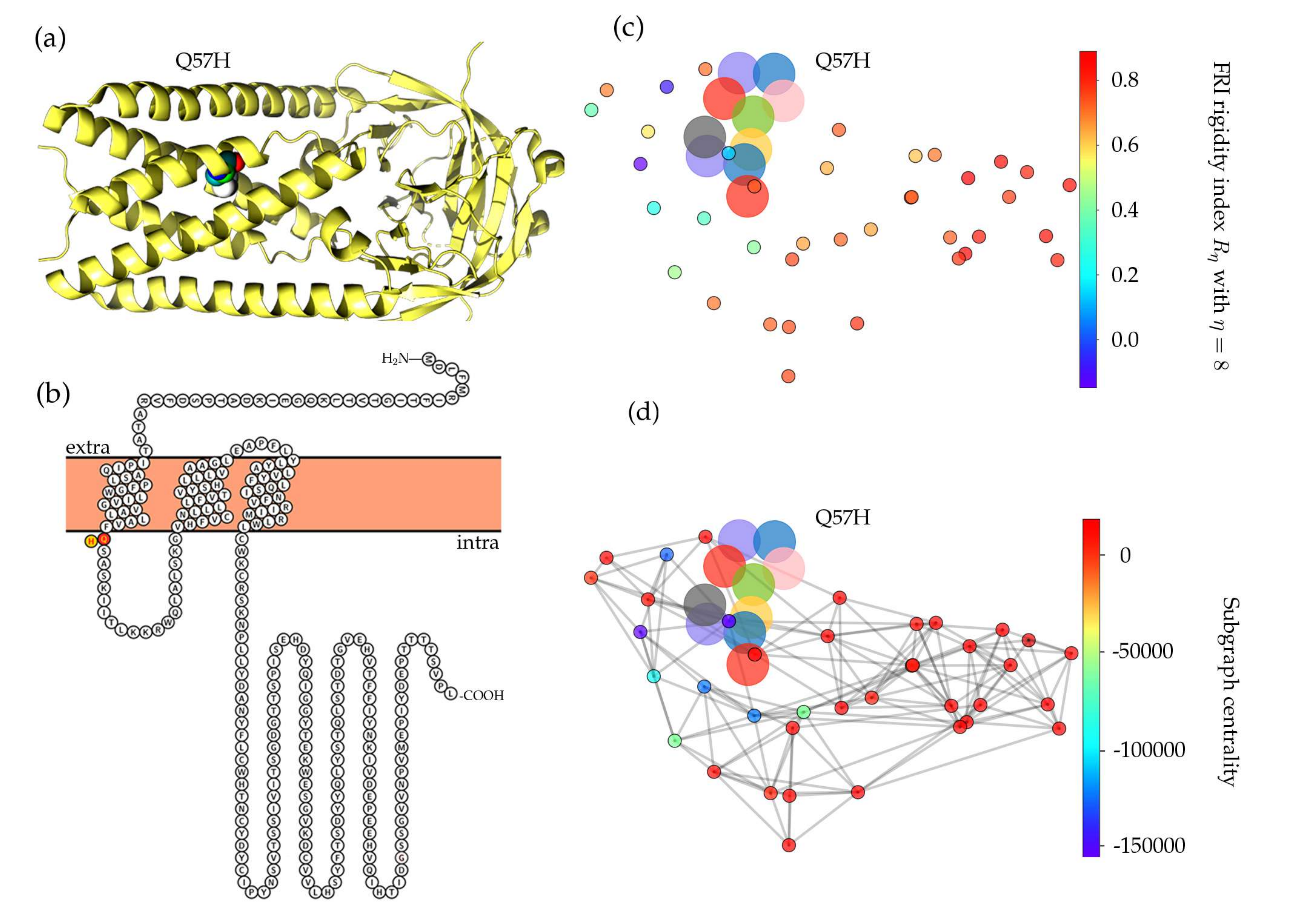}
    \caption{(a) The 3D structure of SARS-CoV-2 ORF3a protein. (b) The visualization of SARS-CoV-2 ORF3a proteoform. The high-frequency mutation 25563G$>$T-(Q57H) on ORF3a is marked in color. The red color represents the wild type and the yellow represents the wild type. (c) The difference of FRI rigidity index between the network with wild type and the network with mutant type. (d) The difference of the subgraph centrality between the network with wild type and the network with mutant type.}
    \label{fig:ORF3a_Proteoform}
\end{figure}

Mutation 25563G$>$T-(Q57H) is on the ORF3a protein. It was first found in  Singapore on February 16, 2020. The case in the United States was detected on February 28, 2020. As of July 14, 2020, 3865 complete genome sequences have the Q57H mutation in the United States, while 6765 isolates carry this type of single mutation in the world. The Q57H mutation changes the amino acid glutamine (Q) with a non-charged polar side chain to the positively charged polar side chain of amino acid histidine (H). \autoref{fig:ORF3a_Align} shows the sequence alignment results in different species. The ORF3a of SARS-CoV-2 and SARS-CoV share a 71.53\% sequence similarity. The red rectangle marks the Q57H mutations with its two neighborhoods. Similarly, one can see that the amino acids nearby position 57 are all conservative in all of 5 species. Moreover, \autoref{fig:ORF3a_Proteoform} (b) visualizes the SARS-CoV-2 ORF3a proteoform, where we use red to mark the wild type amino acid glutamine (Q) and yellow to address the mutant amino acid histidine (H). Spatiotemporally, mutation Q57H  locates at the intramolecular interface and in touch with the membrane, which indicates the special functionality changes that Q57H can induce., and \autoref{fig:ORF3a_Proteoform} (b) is the visualization of ORF3a, which is generated by an online server Protter \cite{omasits2014protter}.

The ORF3a gene of SARS-CoV-2 encodes 275 amino acids with TNF receptor-associated factors (TRAFs), ion channel, and caveolin binding domain. Previous studies have shown that SARS-CoV ORF3a contains a cysteine-rich motif, a tyrosine-based sorting motif, and a diacidic EXD motif, which are the critical domains that participate in the apoptosis and activate the innate immune signaling receptor NLRP3 (NOD-, LRR-, and pyrin domain-containing 3) inflammasome \cite{ren2020orf3a, hassan2020molecular}.  ORF3a of SARS-CoV-2 shares 71.53\% similarity with the ORF3a of SARS-CoV. Similarly, SARS-CoV-2 ORF3a protein is widely expressed in  intracellular and plasma membranes, which induces apoptosis and inflammatory responses in the infected cells and transfected cells \cite{hassan2020molecular}.  
 \autoref{fig:ORF3a_Proteoform} (c) and (d) depict the differences of the FRI rigidity index and the subgraph centrality between the network with wild type and the network with mutant type. We can see that the ORF3a becomes more rigidity after the mutation, which may result in a less flexible mutant for ORF3a to involve in the apoptosis and inflammatory response. 

As illustrated in \autoref{fig:Barplot}, the ratio of the 25563G$>$T-(Q57H) mutation  on ORF3a in each 10-day  period kept increasing once it was introduced to the United States. This tendency indicates that mutation Q57H becomes popular in the viral patients of the United States, which may make the SARS-CoV-2 more infectious. From \autoref{table:Matrix}, we can see that 4827 SNP variants have 25563G$>$T-(Q57H) mutation on ORF3a. Among them, 4818 variants have the [25563G$>$T-(Q57H), 23403A$>$G-(D614G)] co-mutations. As we mentioned above, the S protein plays an important role in viral transmission, indicating that mutation 25563G$>$T-(Q57H) may also be of great importance for viral transmission. The negative folding stability changes of mutation Q57H  in Table \ref{tab:folding stability changes}  reveals that ORF3a becomes unstable following the Q57H mutation, which may harm the function of ORF3a in apoptosis and increase the viral load in the host cell. However, the Q57H mutation locates near TRAF, ion channel, and caveolin binding domain \cite{hassan2020molecular}, which may affect the NLRP3 inflammasome activation. ORF3a activates the innate immune signaling receptor NLRP3 inflammasome via mechanisms such as ion-redistribution and lysosomal disruption, which causes tissue inflammation during respiratory illness and the production of inflammatory cytokines \cite{shah2020novel}. 

\subsubsection{Mutation on the NSP2 protein}

\begin{figure}[ht!]
    \centering
    \includegraphics[width=0.85\textwidth]{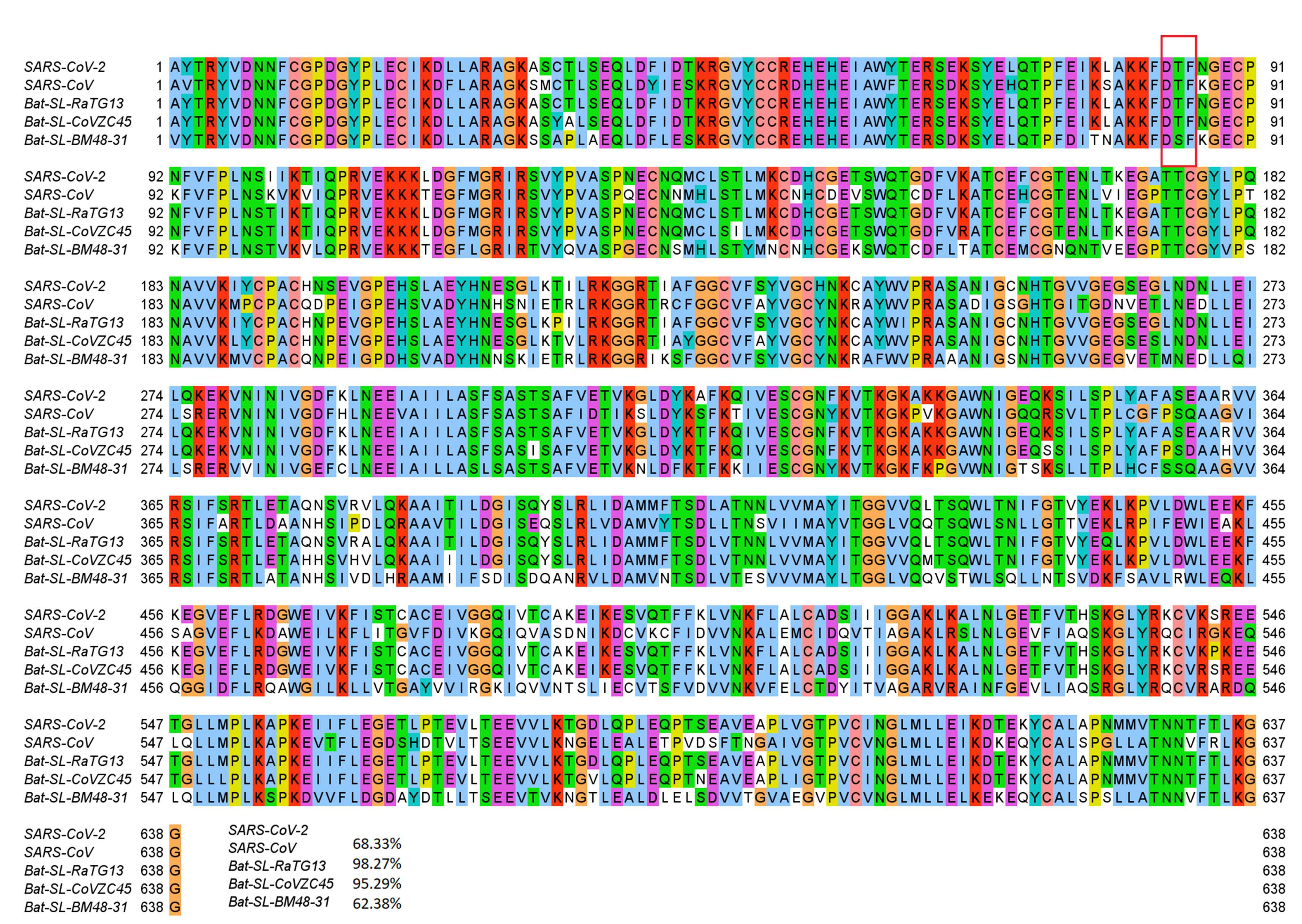}
    \caption{Sequence alignments for the NSP2 of SARS-CoV-2, SARS-CoV, bat coronavirus RaTG13, bat coronavirus CoVZC45, bat coronavirus BM48-31. Detailed numbering is given according to SARS-CoV-2. One high-frequency mutation 1059C$>$T-(T85I) locates on the NSP2 protein. Here, the red rectangle marks the T85I position with its neighborhoods.}
    \label{fig:NSP2_Align}
\end{figure}

\begin{figure}[ht!]
    \centering
    \includegraphics[width=0.9\textwidth]{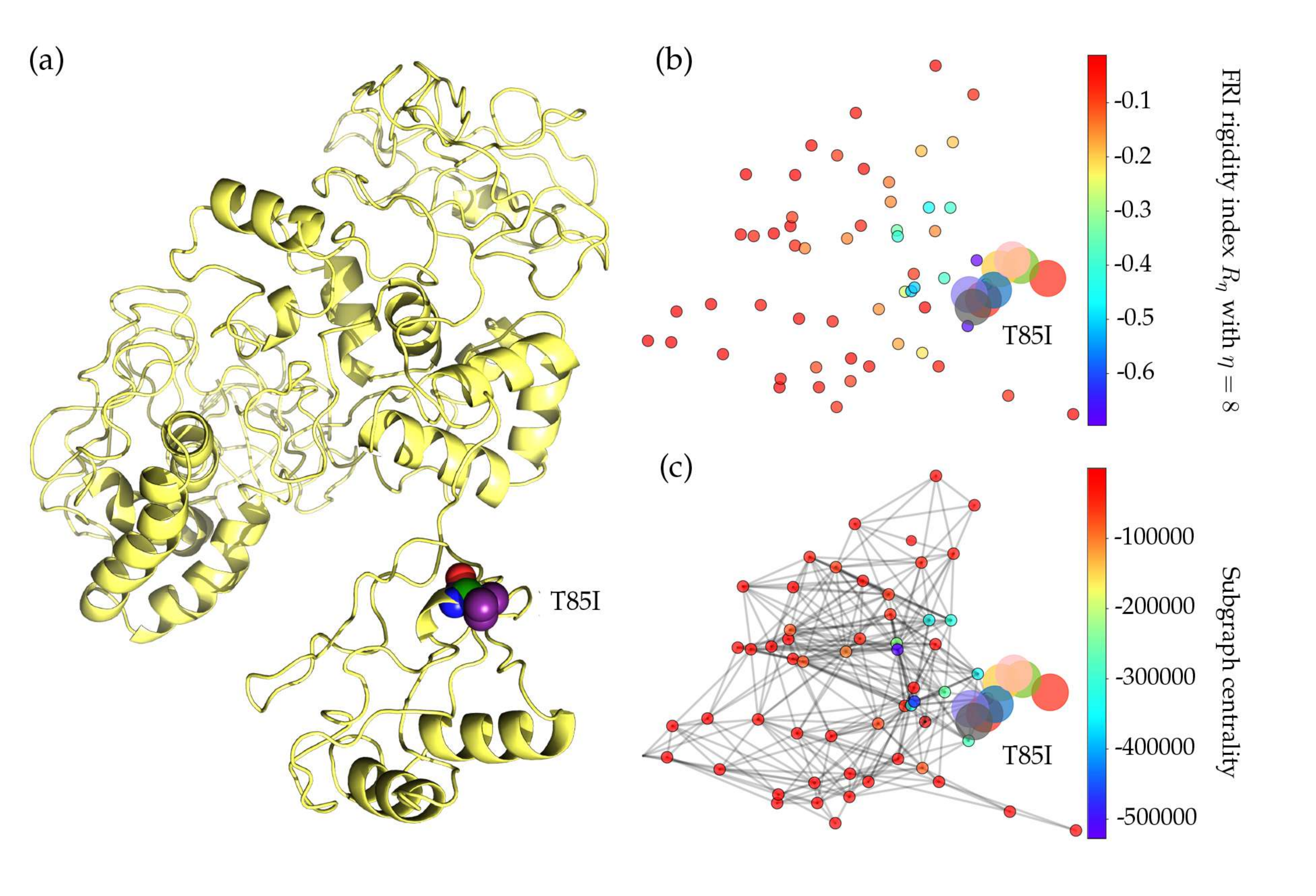}
    \caption{(a) The 3D structure of SARS-CoV-2 NSP2 protein. The mutant residue is marked with color balls. (b) The difference of FRI rigidity index between the network with wild type and the network with mutant type. (c) The difference of the subgraph centrality between the network with wild type and the network with mutant type. }
    \label{fig:NSP2_Pro}
\end{figure}

Mutation 1059C$>$T-(T85I) was first detected in Singapore on February 16, 2020.  On February 29, 2020, the first SNP variants with mutation 1059C$>$T-(T85I) appeared in the United States. As of July 14, 2020, more than half of mutation1059C$>$T-(T85I) counts found worldwide are from the United States. The residue T85 on the NSP2 is polar and non-charged, and it changes to a non-polar residue I85 after the mutation. \autoref{fig:NSP2_Pro} (a) shows the 3D structure of SARS-CoV-2 NSP2. From \autoref{fig:NSP2_Align}, we can see that coronaviral NSP2 is relatively conservative for the first 91 residues. Moreover, the T85 residue with its neighbors is all conservative in the other four SARS-like sequences, indicating this type of mutations may be substantial to coronaviral structures and properties.

NSP2 is also a viral protein that does not attract much research attention. In the SARS-CoV genome, the deletion of NSP2 will only result in a modest reduction in viral titers, which is considered to be a dispensable protein \cite{cornillez2009severe}.   Table \ref{tab:folding stability changes} shows that the folding stability change of T85I is -0.05 kcal/mol.  Although the negative value reveals that T85I may destabilize the structure of NSP2, this small change is negligible. 
 The FRI rigidity change is also minor, as is shown in \autoref{fig:NSP12_Pro} (c), indicating the mutation of T85I on the NSP2 does not change the flexibility of NSP2 too much. However, the growing trend in \autoref{fig:Barplot} still indicates that 1059C$>$T-(T85I) is an infectivity-strengthening mutation, which mainly benefits from the co-mutation with other infectivity-strengthening mutations, such as 23403A$>$G-(DD614) and 25563G$>$T-(Q57H).

\subsubsection{Mutations on the NSP13 protein}
\begin{figure}[ht!]
    \centering
    \includegraphics[width=1\textwidth]{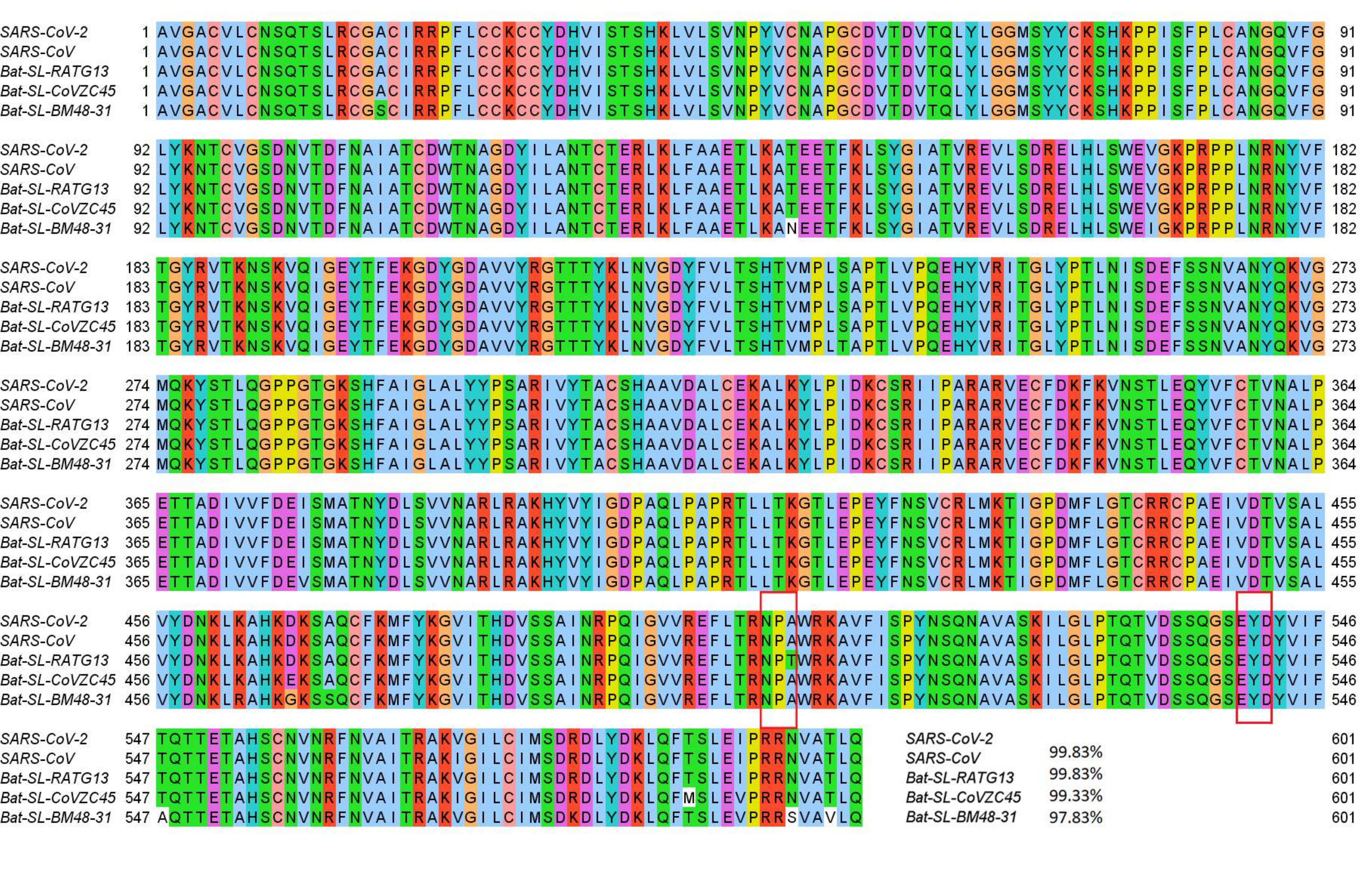}
    \caption{Sequence alignments for the NSP13 protein of SARS-CoV-2, SARS-CoV, bat coronavirus RaTG13, bat coronavirus CoVZC45, bat coronavirus BM48-31. Detailed numbering is given according to SARS-CoV-2. Two high-frequency mutations 17858A$>$G-(Y541C) and 17747C$>$T-(P504L) locate on NSP13. Here, the red rectangles mark the Y541C and P504: mutations with their neighborhoods.}
    \label{fig:NSP13_Align}
\end{figure}

NSP13 of SARS-CoV-2 is a superfamily 1 helicase, which can unwind a double-stranded RNA (dsRNA) or DNA (dsDNA) in the 5' to 3' direction into two single-stranded nucleic acids. Moreover, NSP13 can hydrolyze the deoxyribonucleotide and ribonucleotide triphosphates, which are involved in the ATP-coupling process \cite{adedeji2012mechanism, yuen2020sars}. Furthermore, NPS13 is recruited to the double-membranes vesicles (DMVs) which are mainly found in the cell's perinuclear area during viral infection, indicating that NSP13 is crucial to the viral infection and replication \cite{knoops2008sars}. As illustrated in  \autoref{fig:NSP13_Align}, NSP13 of SARS-CoV-2 shares the most homology with the other 4 species and is one of the most conservative proteins in SARS-CoV-2 genome. Therefore, the existence of two high-frequency mutations on the NSP13 is very unusual. 
 
\begin{figure}[ht!]
    \centering
    \includegraphics[width=0.9\textwidth]{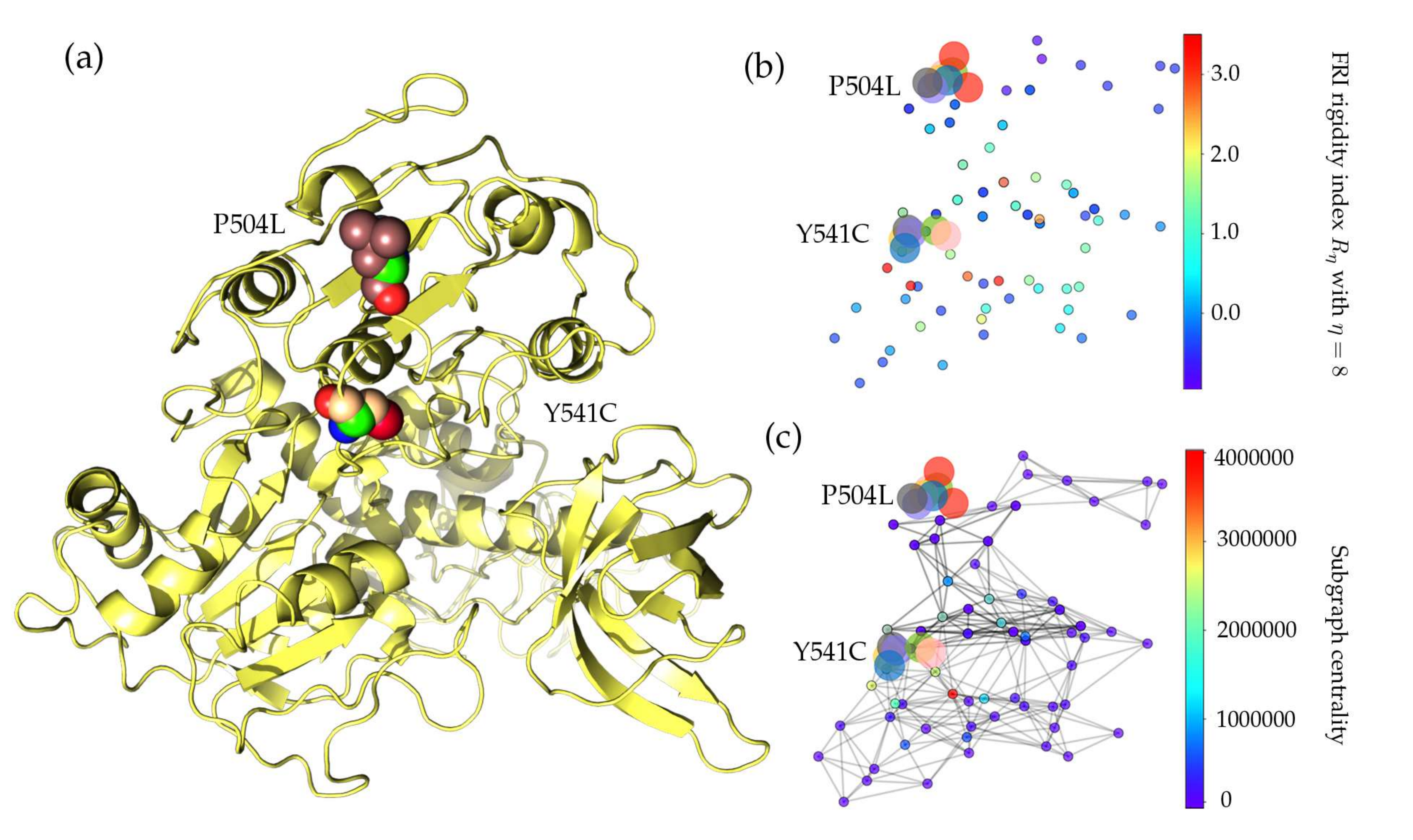}
    \caption{(a) The 3D structure of SARS-CoV-2 NSP13 protein. The mutant residue is marked with color balls. (b) The difference of FRI rigidity index between the network with wild type and the network with mutant type. (c) The difference of the subgraph centrality between the network with wild type and the network with mutant type.}
    \label{fig:NSP13_Pro}
\end{figure}

Similar to 27964C$>$T-(S24L), although 17858A$>$G-(Y541C), 17747C$>$T-(P504L) are in the final list in \autoref{tab:Top 8}, more than 87\% of them were detected in the United States. 
Mutation Y541C changes the amino acid tyrosine (Y) to cysteine (C). \autoref{fig:NSP13_Pro} shows the 3D structure of SARS-CoV-2 NSP13, where the mutant residue is marked with color balls. It is worthwhile to note that tyrosine has an aromatic side chain, while cysteine only has a polar non-charged side chain, indicating that the 3D structure of NSP13 will be incredibly affected. Moreover, the other single mutation at the residue position 504 will not change the structure of NSP13 very much since both the mutant and wild-type residues are non-polar aliphatic residues. 

 From \autoref{fig:Barplot}, one can note that both mutations on the NSP13 have the same trajectory  of the mutation ratios over time. Once mutations 17858A$>$G-(Y541C) and 17747C$>$T-(P504L) were first found in the United States, they had a rapid increase in the first two weeks. However, these two mutations do not benefit SARS-CoV-2. In early March, the ratio of both mutations start to decrease and approach zero after May 19, 2020, suggesting that mutations 17858A$>$G-(Y541C) and 17747C$>$T-(P504L) may hinder the transmission of SARS-CoV-2. It is interesting to note that in \autoref{table:Matrix}, 17858A$>$G-(Y541C) and 17747C$>$T-(P504L) never show up with 23403A$>$G-(D614G) in more than a thousand SNP variant, which provides a side evidence that these two mutations may inhibit the contagiousness of SARS-CoV-2.

\autoref{tab:folding stability changes} shows that both high-frequency mutations Y541C and P504L have negative folding stability changes, which will destabilize the structure of NSP13. As stated earlier, mutations 17858A$>$G-(Y541C) and 17747C$>$T-(P504L) happen simultaneously after analyzing 24715 genome sequences, which means the folding stability changes on the NSP13 are superimposed by two simultaneously occurred mutations. This also explains the same decreasing tendency in \autoref{fig:Barplot} after early March. Based on the protein-specific analysis mentioned above, we can deduce that mutations Y541C and P504L prevent SARS-CoV-2 from efficiently interacting with host interferon signaling molecules and impede the NSP13 from efficacious participation in the replication/transcription process.  \autoref{fig:NSP12_Pro} (b) shows the difference of the FRI rigidity index between the network with wild type and the network with the mutant type. One mutation (17747C$>$T-(P504L)) does not affect the rigidity much whereas the other mutation (17858A$>$G-(Y541C)) leads to a significant decrease in the  NSP12 rigidity, which may make NSP13 not as robust as before to involve in the viral infection and replication process. 

\subsubsection{Mutations on the ORF8 protein}

\begin{figure}[ht!]
    \centering
    \includegraphics[width=1\textwidth]{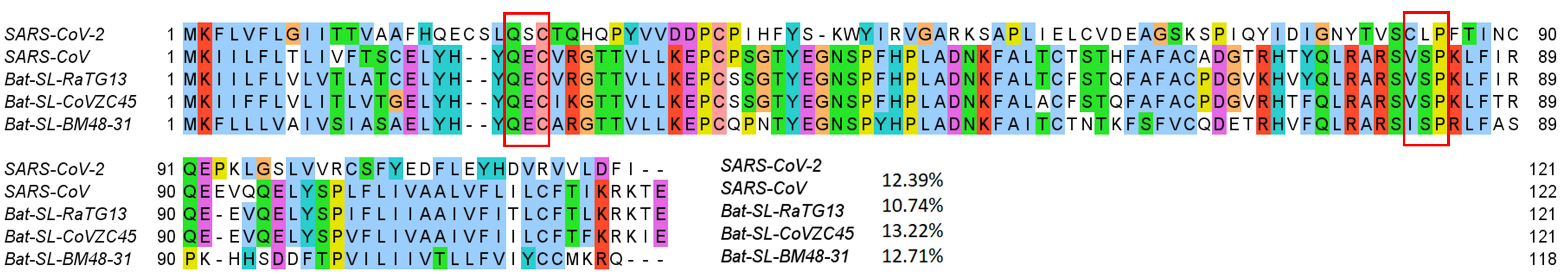}
    \caption{Sequence alignments for the ORF8 protein of SARS-CoV-2, SARS-CoV, bat coronavirus RaTG13, bat coronavirus CoVZC45, bat coronavirus BM48-31. Detailed numbering is given according to SARS-CoV-2. Two high-frequency mutations 28144T$>$C-(L84S) and 27964C$>$T-(S24L) locate on the ORF8. Here, the red rectangles mark the S24L and L84S mutations with their neighborhoods.}
    \label{fig:ORF8_Align}
\end{figure}

The ORF8 protein also has two high-frequency mutations,  28144T$>$C-(L84S) and 27964C$>$T-(S24L). Notably, although 27964C$>$T-(S24L) has the lowest frequency in the top 8 missense mutations, more than 94.1\% mutation 27964C$>$T-(S24L) worldwide were found in the United States. Moreover, the first confirmed case with 27964C$>$T-(S24L) was discovered on February 20, 2020, in the United States, suggesting that S24L  initially happened in the US. Another mutation on ORF8 is 28144T$>$C-(L84S), the number of sequences with the L84S mutation in the United States is 1437, which accounts for more than 50\% proportion of the isolates in the world. It is interesting to address that these two high-frequency mutations S24L and L84S mutate reversibly. The amino acid serine (S) has a non-charged polar side chain, while the leucine (L) has a non-polar aliphatic residue. \autoref{fig:ORF8_Align} illustrates the sequence alignment of SARS-CoV-2 ORF8 with the other 4  species. The SARS-CoV-2 ORF8 shares a really low similarity among all the other four SARS-like species. SARS-CoV, Bat coronavirus RaTG13, Bat coronavirus CoVZC45, and Bat coronavirus BM48-31 have the same residues at positions 24 and 84. Nonetheless, SARS-CoV-2 ORF8 owns different types of residues. Here, we would like to address that although the ORF8 of SARS-CoV-2 at position 84 has a different residue compared to the other 4 species, it mutates back to S in 1434 isolates in the United States.

\begin{figure}[ht!]
    \centering
		    \includegraphics[width=0.75\textwidth]{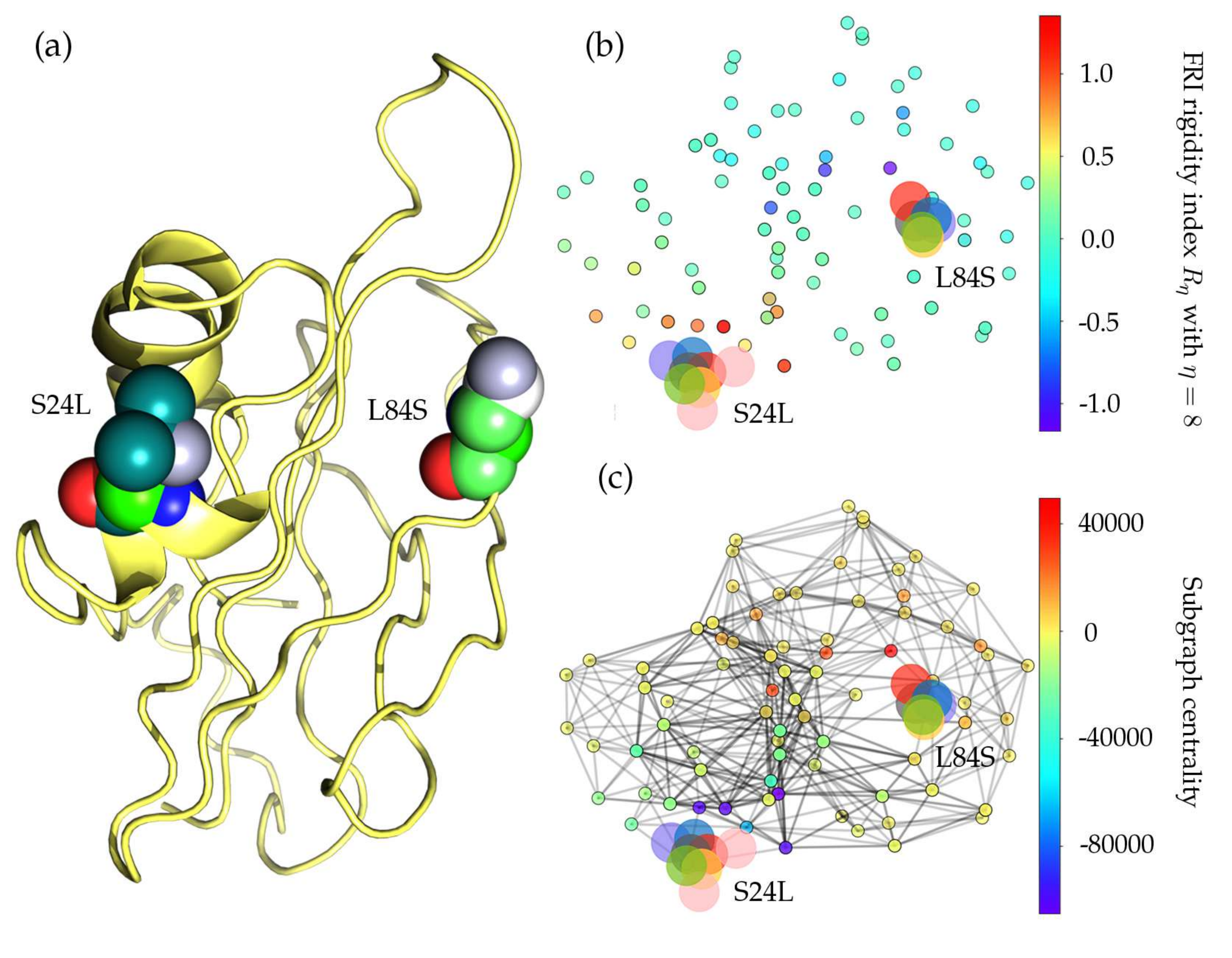}
    \caption{(a) The 3D structure of SARS-CoV-2 ORF8 protein. The mutant residue is marked with color balls. (b) The difference of FRI rigidity index between the network with wild type and the network with mutant type. (c) The difference of the subgraph centrality between the network with wild type and the network with mutant type.}
    \label{fig:ORF8_Pro}
\end{figure}

\autoref{fig:ORF8_Pro} (a) shows the 3D structure of SARS-CoV-2 ORF8. ORF8 protein of SARS-CoV-2 shares the least homology with SARS-CoV among all viral proteins, which mediates the immune evasion by interacting with major histocompatibility complex molecules class I (MCH-I) and down-regulating the surface expression of MHC-I on various cells \cite{zhang2020orf8}. Once the ``outer" invaders or antigens (i.e., viruses) enter the host cell, the MCH-I will bind to them and bring them to the surface of the infected cell. Then, the T-cell receptors (TCRs) that are expressed by cytotoxic T cells (TCLs) can recognize this unique signal presented by MHC-I-peptide complex and directly eradicate the virus-infected cells, which is an effective and efficient anti-viral immune response \cite{ni2020detection}. However, in  ORF8-expressing cells, the MHC-I protein becomes the major target for the lysosomal degradation mediated by bafilomycin A1 (Baf-A1), resulting in the limited expression of MHC-I. Therefore,  ORF8 protein has an effect on downregulating MHC-I, which will favor SARS-CoV-2 to evade immune surveillance by hindering the presentation of antigen regulated by MHC-I \cite{zhang2020orf8}. 

In \autoref{fig:Barplot}, the ratio of S24L per 10 days is slightly going up before June 2, 2020. In the next ten days' period, the sharp increase and sharp drop in the ratio are due to the limited complete genome sequences submitted to GISAID in June. However, the overall upward trend of the S24L ratio over time reveals that S24L may enhance SARS-CoV-2's ability to spread. In contrast, the time evolution plot shows that the ratio of mutation 28144T$>$C-(L84S) goes up before the beginning of March, and then the ratio goes down and approach zero after May 23, 2020. Due to the small number of sequence data, we can say that the ratio of L84S has a decreasing tendency. 

As discussed earlier,  the female patients with S24L mutation on ORF8 account for a large proportion, which indicates that the S24L is most likely to happen in the female population in the United States.

\autoref{tab:folding stability changes}  shows that the folding stability change of 28144T$>$C-(L84S) is -0.99 kcal/mol, indicating that ORF8 becomes unstable.   \autoref{fig:ORF8_Pro} (b) shows that the ORF8 becomes slightly less rigidity after both L84S and S24L mutations. To be noted, the rigidity changes induced by S24L is less than the L84S.  Based on the function of ORF8 that involved in the immune response, we deduce that L84S may disfavor SARS-CoV-2 and  favor the host immune surveillance to decrease the viral load in the human cells, which provides an explanation that the ratio of L84S in \autoref{fig:Barplot} keeps decreasing. Meanwhile, the positive folding stability change of 27964C$>$T-(S24L) lists in \autoref{tab:folding stability changes} reveals that this type of mutation may enhance the function of ORF8. Therefore, the MHC-I will be inhibited more, and the eradication of SARS-CoV-2 in vivo will be hindered. This explains the increasing trend in the ratio of S24L. Notably, after analyzing 28726 complete genome sequences, none of them have mutations 28144T$>$C-(L84S) and 27964C$>$T-(S24L) happened simultaneously.

\subsection{Infectivity analysis }\label{sec:Infectivity}
As mentioned above, SARS-CoV-2 enters the host cell by the interaction of the S protein and ACE2. The viral S protein is primed by TMPRSS2 to entail its cleavage at two potential sites, Arg685/Ser686 and Arg815/Ser816 \cite{hoffmann2020sars}. Based on the 7823 SNP variants' information in the United States, we found 264 non-degenerate single mutations on the spike protein. Among them, 32 single mutations are detected on the receptor-binding domain (RBD). Moreover, 7 single mutations occurred on the receptor-binding motif (RBM), the region that directly connects with the ACE2.  In this section, we separate  7823 SNP variants in the United States into four clusters and calculate the mutation-induced binding affinity changes of S protein RBD and ACE2 in each cluster, which will help us understand the potential transmission tendency induced by the mutations on the S protein RBD. The binding affinity change induced by single mutation $\Delta\Delta G = \Delta G_W-\Delta G_M$  is defined as the subtraction of the binding affinity of the mutant type ($\Delta G_M$) from the binding affinity of the wild type ($\Delta G_W$). Furthermore, the positive binding affinity change of a single mutation means that the mutation can enhance the binding affinity of the S protein RBD and ACE2 and make SARS-CoV-2 more infectious.

\begin{figure}[ht!] 
    \centering
    \includegraphics[width=0.8\textwidth]{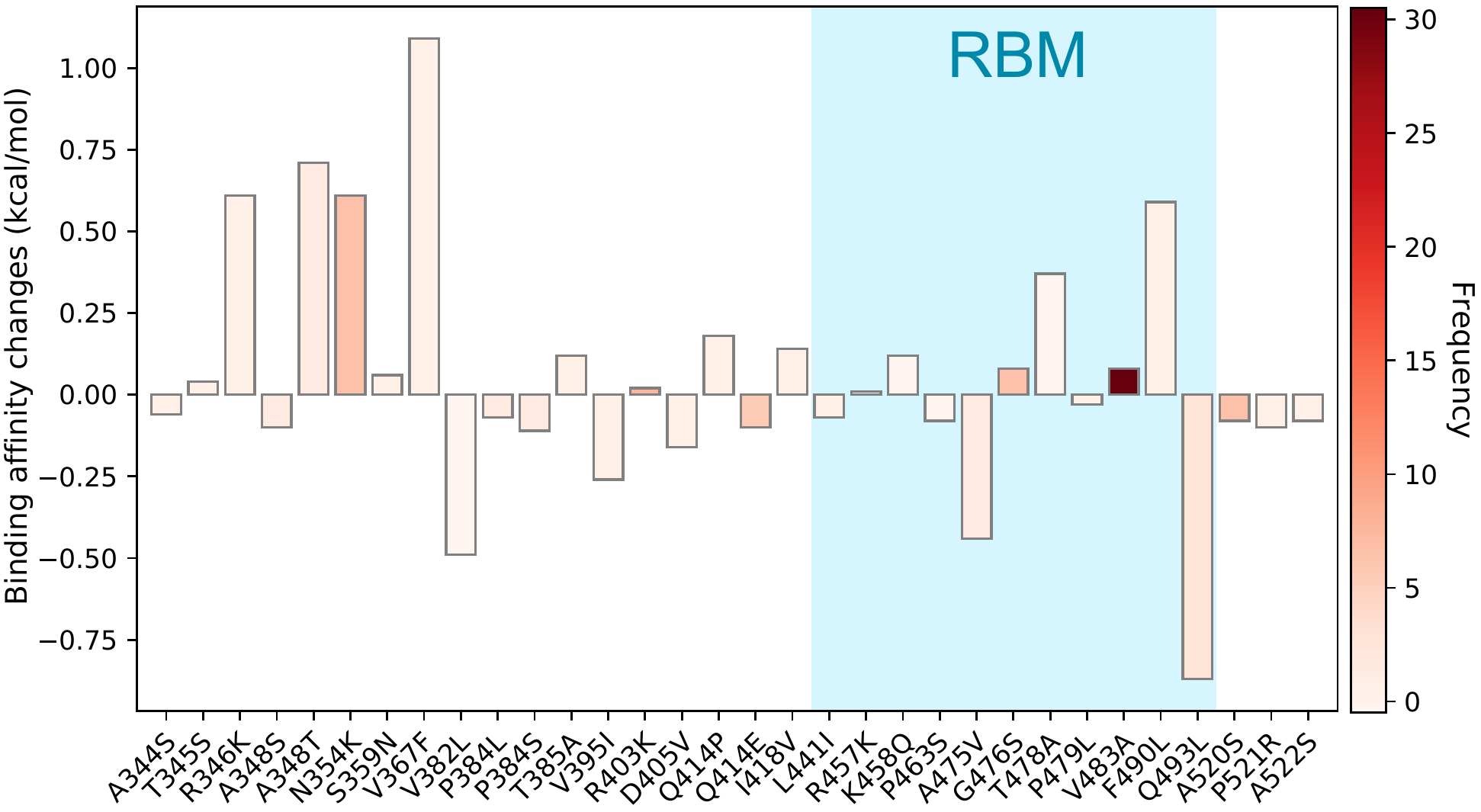}
    \caption{Overall binding affinity changes $\Delta\Delta G$ (kcal/mol) on the receptor-binding domain (RBD). The blue color region marks the binding affinity changes on the receptor-binding motif (RBM). The height of each bar indicates the predicted $\Delta\Delta G$. The color indicates the occurrence frequency in the GISAID genome dataset.}
    \label{fig:usMutation}
\end{figure}

\autoref{fig:usMutation} illustrates overall binding affinity changes $\Delta\Delta G$ (kcal/mol) induced by 32 single mutations on SARS-CoV-2 S protein RBD. The color bar on the left-hand side of the figure represents the mutation frequency. We can see that 50\% single mutations have positive binding changes (16 out of 32). Moreover, the frequency of mutations with positive binding affinity changes is higher than those with negative binding affinity changes, suggesting that SARS-CoV-2 is more likely to be infectious. Notably, the mutation 23010T$>$C-(V483A) has the highest frequency (30) localized on the RBM has the positive binding affinity change, which indicates that V483A is prevalent in COVID-19 patients' in the United States has a potential capacity to enhance the infectivity of SARS-CoV-2. However, mutations that locate away from the RBM will also have a crucial impact on the infectivity \cite{chen2020mutations}. Although away from the RBM, the relatively high frequency and positive binding affinity changes of N354K, R403K, and G467S indicate more attention  should be paid to them in the future. Additionally, an interesting finding is that the mutations occurred at the same residue position such as A348S and A348T, P384L and P384S, and Q414E and Q414P always have similar binding affinity changes.

\begin{figure}[ht!]
	\centering
		\includegraphics[width=1\linewidth]{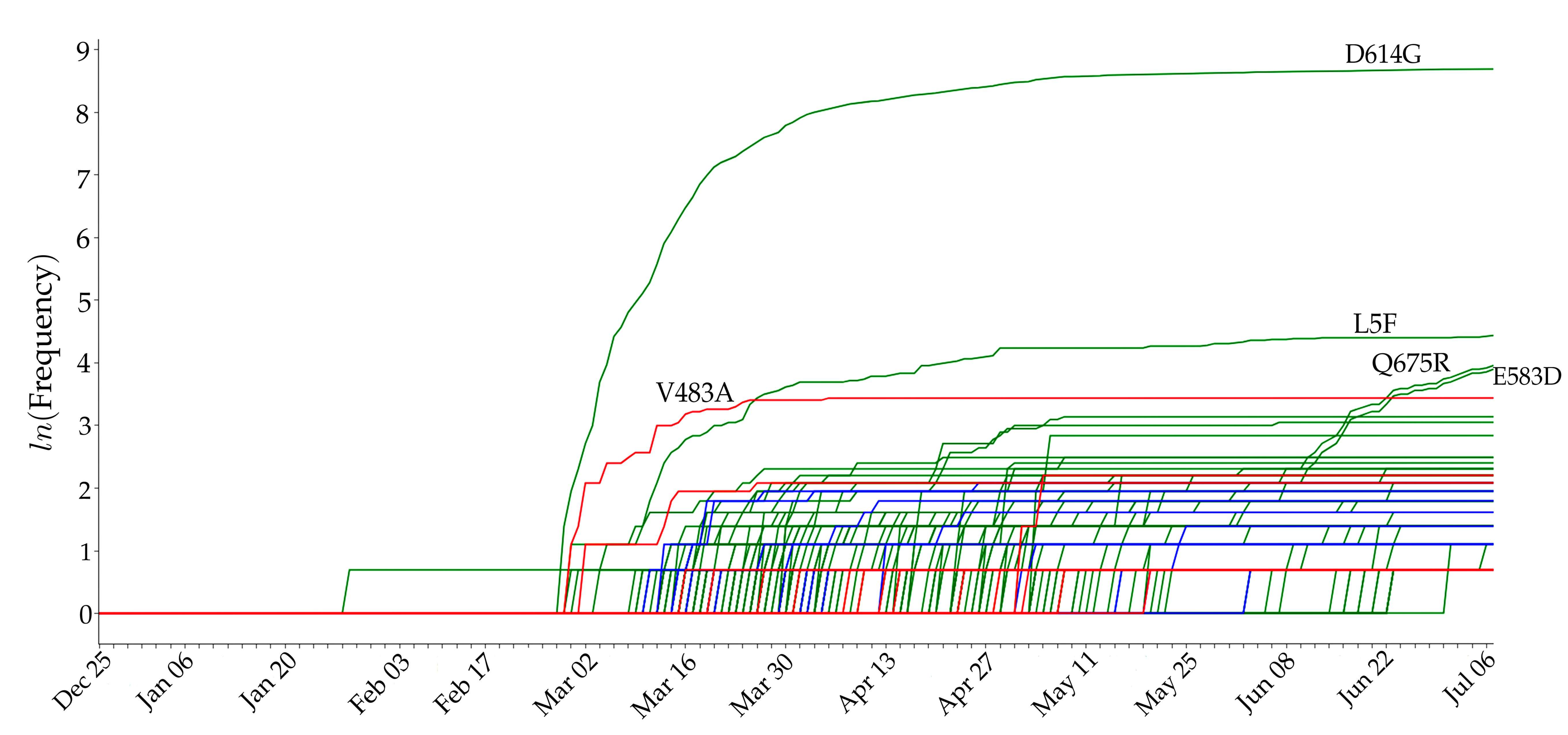}
	\caption{The time evolution  of 264  SARS-CoV-2 S protein mutations. The red lines represent the RBD mutations that strengthen the infectivity of SARS-CoV-2 (i.e., $\Delta\Delta G$ is positive), the blue lines represent the RBD mutations that weaken the infectivity of SARS-CoV-2 (i.e., $\Delta\Delta G$ is negative), and the green lines are for S protein mutations that away from the RBD. 
	The mutation with the highest frequency is D614G.  
	}
	\label{fig:lineplot_US}
\end{figure}

 \autoref{fig:lineplot_US} illustrates the time evolution trajectories of 274 single mutations on SARS-CoV-2 S protein RBD. The red line shows the mutations that have positive binding affinity changes and the blue lines represent the mutations that have negative binding affinity changes. 
Here, we hypothesize that single mutations on S protein RBD with positive binding affinity changes will enhance the viral transmission since natural selection favors them. One can see that the red lines gradually outpace the blue lines as time progresses, suggesting that our hypothesis is correct. 

Additionally, green lines indicate the evolutions of mutations that locate away from the RBD. The mutation that has the highest frequency is D614G, which was reported to enhance SARS-CoV-2 infectivity \cite{korber2020spike, zhang2020d614g}. The trajectories of the other two high frequency S protein mutations (Q675R and E583D) indicate that they are co-mutations with infectivity-enhancing S protein mutations, such as D614G.  We found that the other high frequency S protein mutation L5F is independent of mutation D614G.

Based on the genotyping results, we separate 7823 SNP variants from the United States into four clusters. \autoref{table:US} shows the mutation distribution of four clusters with number of samples ($N_{\rm NS}$) and total single mutation counts ($N_{\rm TF}$) in 20 states. Accordingly, a more specific analysis of the relationship between the sign of the binding affinity changes and the transmission ability is discussed below.

\subsubsection{Cluster A infectivity}
\begin{figure}[ht!]

    \centering
    \includegraphics[width=1\textwidth]{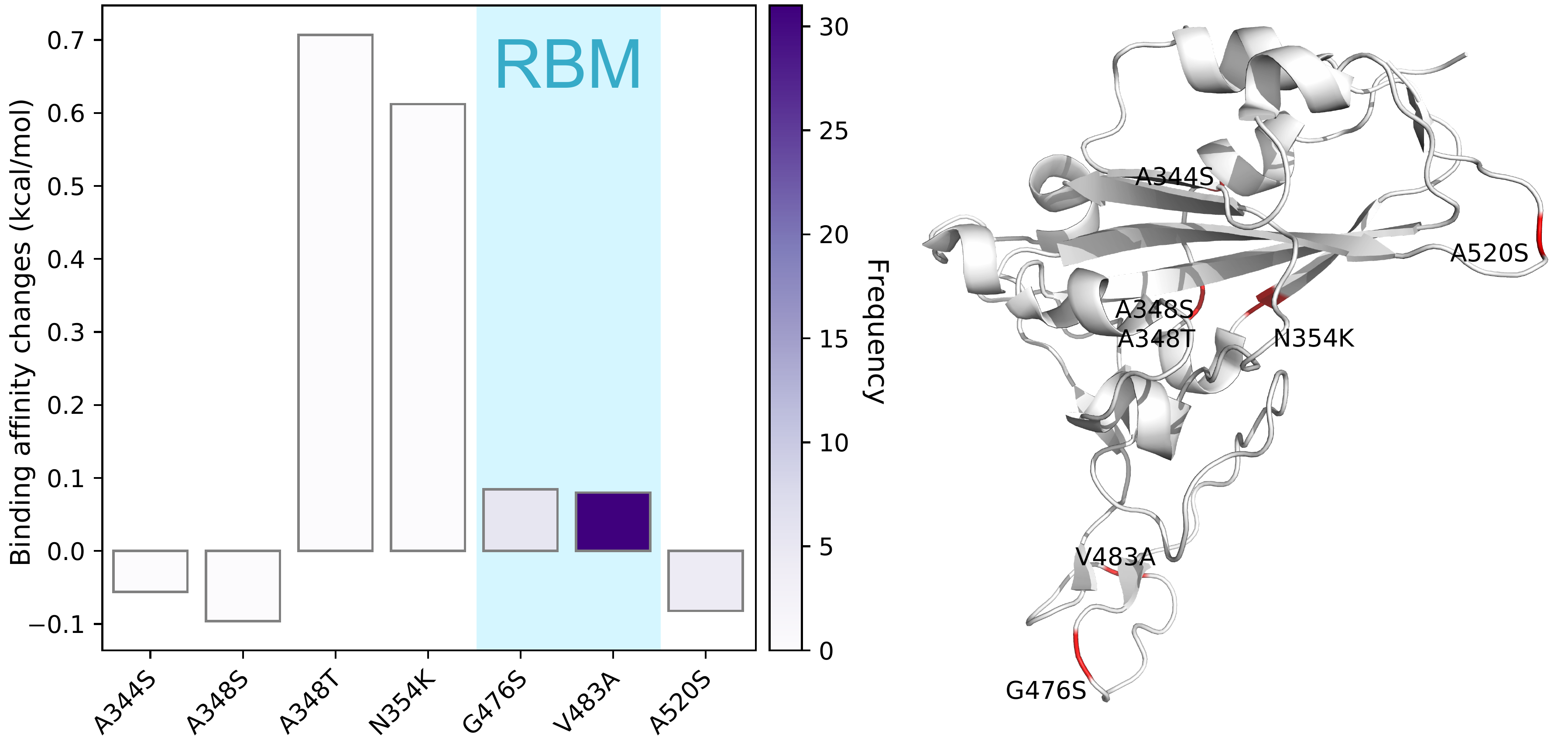}
    \caption{Cluster A. Left: binding affinity changes $\Delta\Delta G$  (kcal/mol) induced by mutations in Cluster V. Right: mutations on the SARS-CoV-2 S protein RBD.}
    \label{fig:ClusterA}
\end{figure}

 \autoref{fig:ClusterA} depicts the binding affinity changes of mutations in Cluster A. Total seven single mutations are found in Cluster A. Four of them have the positive binding affinity changes, while the other three mutations induced the negative binding affinity changes. Therefore, the mutations in Cluster B will strengthen the infectivity of SARS-CoV-2 in general. The V483A mutation is localized on the RBM with the highest frequency, indicating that V483A may favor SARS-CoV-2 by natural selection and cause SARS-CoV-2 more infectious. Although A348T and N354K have relatively low frequencies due to the limited number of genome samples, their high binding affinity changes lead to a more contagious SARS-CoV-2 substrain. It is worth noting that from \autoref{table:US}, mutations in Cluster A are involved in all of the 20 states except for LA. However, DC, MA, and TX each only have one SARS-CoV-2 isolate related to Cluster A. Therefore, LA, DC, MA, and TX are not contributed to the infectivity-strengthening mutations in Cluster A. 

\subsubsection{Cluster B  infectivity}

\begin{figure}[ht!]
    \centering
    \includegraphics[width=1\textwidth]{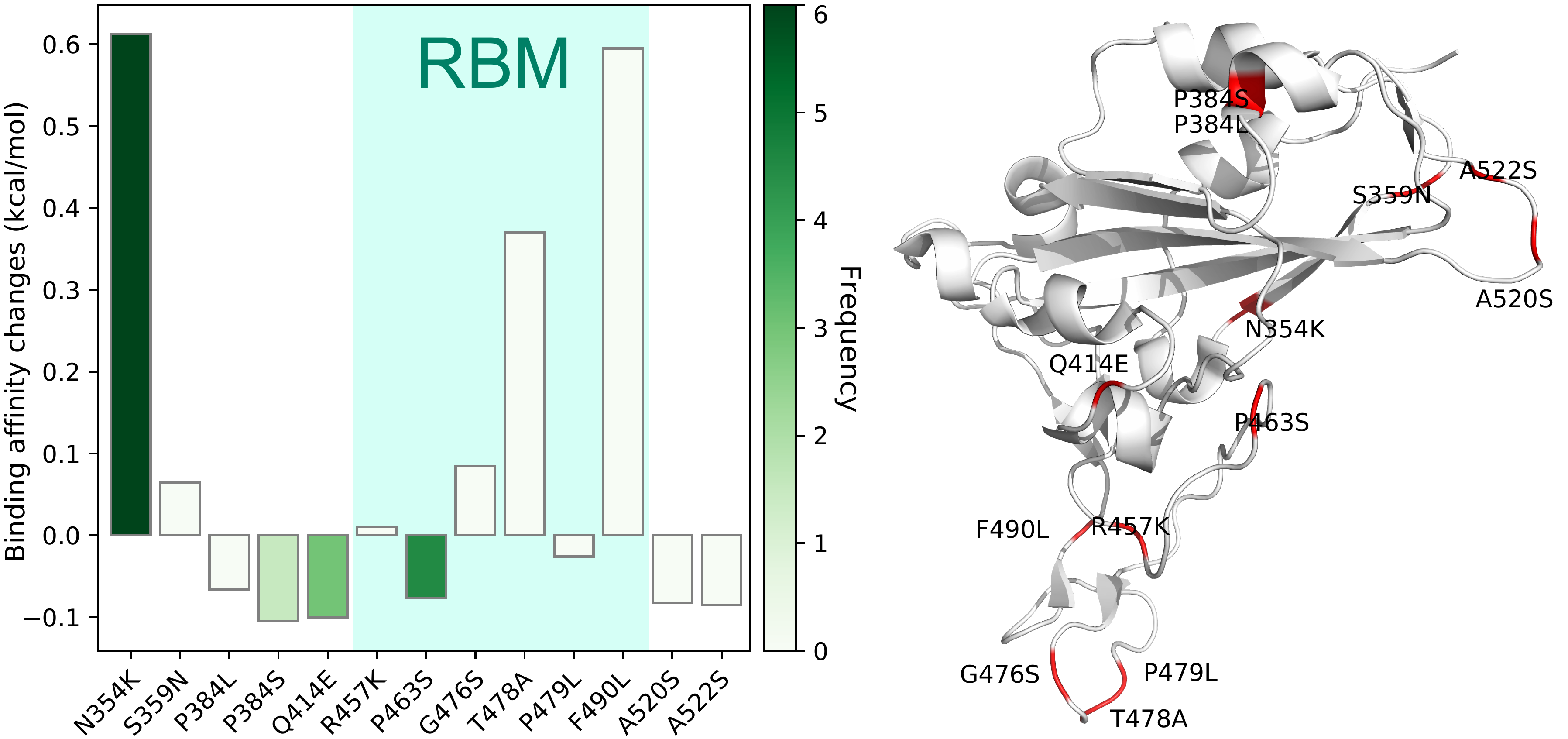}
    \caption{Cluster B. Left: binding affinity changes $\Delta\Delta G$  (kcal/mol)  induced by mutations in Cluster B. Right: mutations on the SARS-CoV-2 S protein RBD.}
    \label{fig:ClusterB}
\end{figure}

\autoref{fig:ClusterB} describes the binding affinity changes of mutations in Cluster B. There are thirteen single mutations on the S protein RBD and six single mutations on the RBM. Although the number of single mutations with positive binding affinity changes is less than those with negative binding affinity changes, the high frequency of N354K slightly enhances the infectivity of SARS-CoV-2. We can notice that all of the states in the \autoref{table:US} are associated with Cluster B. Additionally, a large proportion of single mutations in Cluster B are in CA, LA, MN, MI, NY, and WA.

\subsubsection{Cluster C  infectivity}
\begin{figure}[ht!]
    \centering
    \includegraphics[width=1\textwidth]{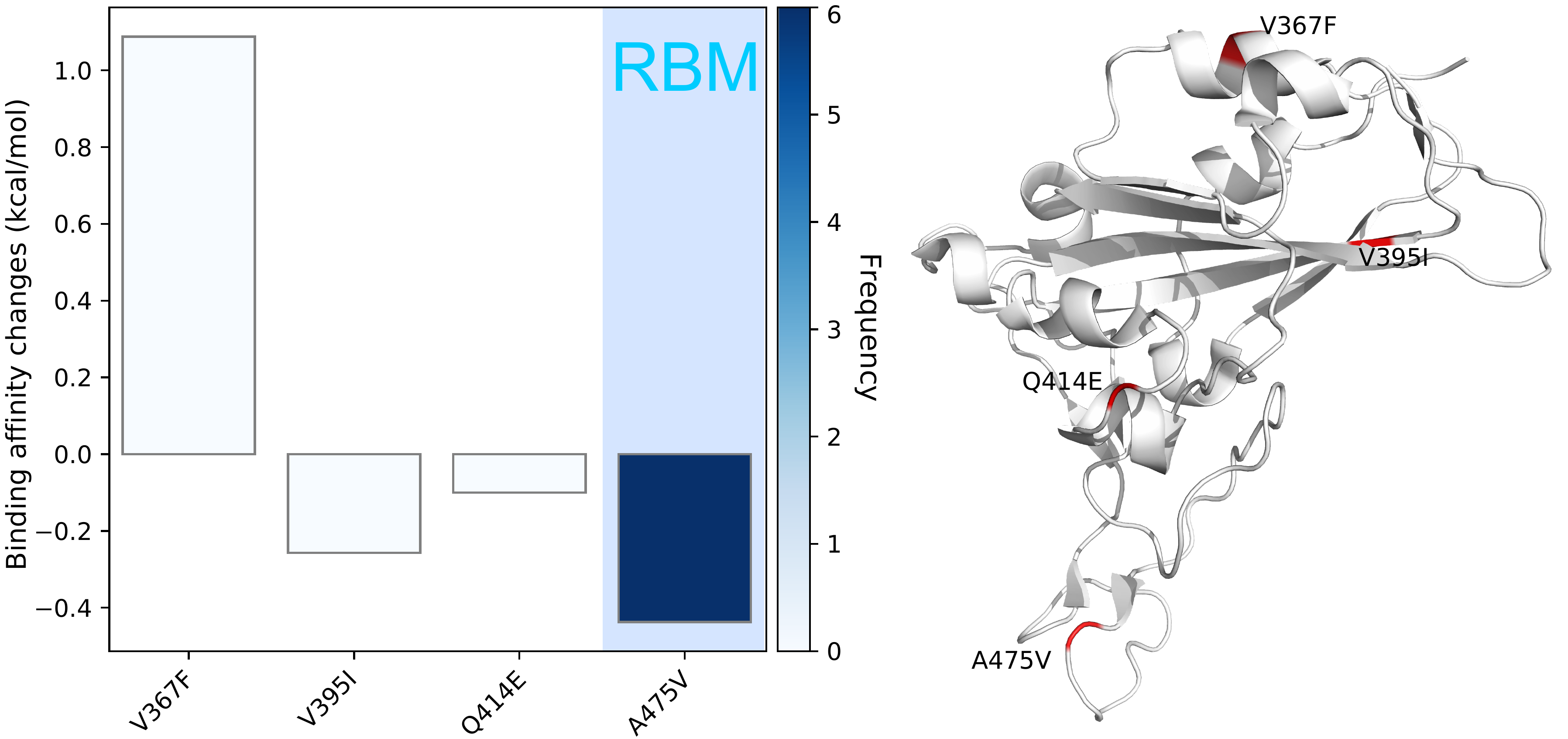}
    \caption{Cluster C. Left: binding affinity changes $\Delta\Delta G$  (kcal/mol)  induced by mutations in Cluster C. Right: mutations on the SARS-CoV-2 S protein RBD.}
    \label{fig:ClusterC}
\end{figure}

\autoref{fig:ClusterC} describes the binding affinity changes in Cluster C. This is the only cluster that has more infectivity-weakening RBD mutations. To be noted, A475V on the RBM has a negative binding affinity change with a relatively high frequency compared to the other 3 mutations in Cluster C. However, only four mutations are detected in the SNP variants that belong to Cluster C, indicating that Cluster C has a limited proportion of the infected patients in the United States. Our conjecture is confirmed  by  \autoref{table:US}. Although 20 states are involved in Cluster C except for NM, fewer samples related to the mutations in Cluster C are found compared to the other three clusters. This means the infectivity-weakening mutations do not rapidly spread across the United States. However, CA and WI have slightly more significant number of samples, indicating that the recent situation in these states is better than the other states.

\subsubsection{Cluster D  infectivity}
\begin{figure}[ht!]
    \centering
    \includegraphics[width=1\textwidth]{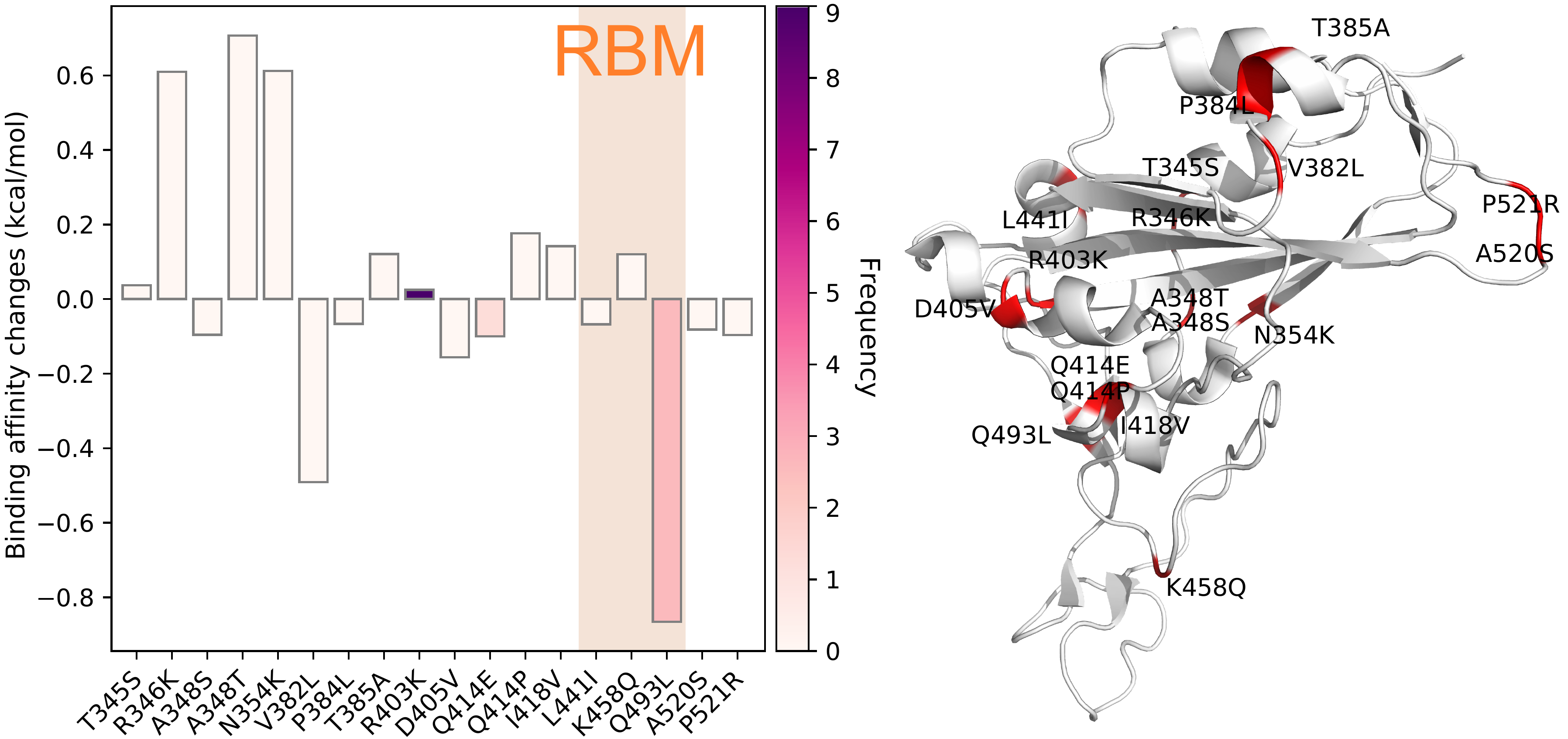}
    \caption{Cluster D. Left: binding affinity changes $\Delta\Delta G$  (kcal/mol)  induced by mutations in Cluster D. Right: mutations on the SARS-CoV-2 S protein RBD.}
    \label{fig:ClusterD}
\end{figure}

The binding affinity changes of RDB mutations in Cluster D are illustrated in  \autoref{fig:ClusterD}. Eighteen  different single mutations are classified in Cluster D. Among them, nine mutations have positive binding affinity changes and relatively higher frequencies, showing that overall the mutations in Cluster D can enhance the transmission capacity of SARS-CoV-2. From \autoref{table:US}, we can see that R346K, A348T, and N354K have relatively high binding affinity changes. In addition, R403K has the highest frequency among the nine infectivity-strengthen mutations. From \autoref{table:US}, all of these 20 states have a large proportion in Cluster D, suggesting that the infectivity-strengthen mutations are widely spreading in the United States, especially in  CA, MN, NY, WA, and WI.

Finally, since the infectivity-strengthening D614G mutation is associated  with all clusters and essentially all the US genome isolates, it may be quite reasonable to say all of the US SARS-CoV-2 substrains become more infectious compared with the original genome collected on December 24, 2019 in China.     

\section{Methods}

\subsection{Data collection and pre-processing}

On January 5, 2020, the complete genome sequence of SARS-CoV-2 was first released on the GenBank (Access number: NC\_045512.2) submitted Zhang's group at Fudan University \cite{wu2020new}. Since then, there has been a rapid accumulation of SARS-CoV-2 genome sequences. In this work,   24,715 complete genome sequences with high coverage of SARS-CoV-2 strains from the infected individuals in the world were downloaded from the GISAID database \cite{shu2017gisaid} (\url{https://www.gisaid.org/}) as of July 10, 2020. All the incomplete records and those without the exact submission date in GISAID  were not considered. To rearrange the complete genome sequences according to the reference SARS-CoV-2 genome, multiple sequence alignment (MSA) is carried out by using Clustal Omega \cite{sievers2014clustal} with default parameters. 

The amino acid sequence of NSP2, NSP12, NPS13, Spike protein, ORF3a, ORF8 were downloaded from the GenBank \cite{benson2009genbank}. The three-dimensional (3D) structures of NSP12, spike protein, and ORF3a used in this work were extracted from the Protein Data Bank (\url{https://www.rcsb.org/}), denoted as 7BTF, 6VYB, and 6XDC, respectively. The 3D structures of NSP2, ORF8, and NSP13 were generated by I-TASSER model \cite{yang2015tasser}. The 3D structure graph is created by using PyMOL \cite{delano2002pymol}.

\subsection{Single nucleotide polymorphism genotyping}

Single nucleotide polymorphism (SNP) genotyping measures the genetic variations between different members of a species. Establishing the SNP genotyping method to the investigation of the genotype changes during the transmission and evolution of SARS-CoV-2 is of great importance \cite{yin2020genotyping, wang2020decoding}. By analyzing the rearranged genome sequences, SNP profiles, which record all of the SNP positions in teams of the nucleotide changes and their corresponding positions, can be constructed. The SNP profiles of a given SARS-CoV-2  genome isolated from a COVID-19 patient capture all the differences from a complete reference genome sequence and can be considered as the genotype of the individual SARS-CoV-2. 

\subsection{Distance of SNP variants}

In this work, we use the Jaccard distance to measure the similarity between SNP variants and compare the difference between the SNP variant profiles of SARS-CoV-2 genomes. 

The Jaccard similarity coefficient is defined as the intersection size divided by the union of two sets $A$ and $ B$ \cite{levandowsky1971distance}:
\begin{equation}
    J(A,B) = \frac{|A\cap B|}{|A\cup B|} = \frac{|A\cap B|}{|A|+|B|-|A\cap B|}.
\end{equation}
The Jaccard distance of two sets $A$ and $B$ is scored as the difference between one and the Jaccard similarity coefficient and  is a metric on the collection of all finite sets:
\begin{equation}
    d_{J}(A,B) = 1 - J(A, B) = \frac{|A\cup B| - |A\cap B|}{|A\cup B|}.
\end{equation}
Therefore, the genetic distance of two genomes corresponds to the Jaccard distance of their SNP variants.

 In principle, the Jaccard distance of SNP variants takes account of the ordering of SNP positions, i.e., transmission trajectory, when an appropriate reference sample is selected. However, one may fail to identify the infection pathways from the mutual Jaccard distances of multiple samples.  In this case,  the dates of the sample collection provide key information. Additionally,  clustering techniques, such as $k$-means, UMAP, and t-distributed stochastic neighbor embedding (t-SNE), enable us to characterize the spread of COVID-19 onto the communities.

\subsection{$K$-means clustering}

$K$-means clustering aims at partitioning a given data set $X=\{x_1, x_2, \cdots, x_n, \cdots, x_N\}, x_n \in \mathbb{R}^d$ into $k$ clusters $\{C_1, C_2, \cdots, C_k\}, k \le N$ such that the specific clustering criteria are optimized. The standard $K$-means clustering algorithm picks $k$ points as cluster centers randomly at beginning and  separates each data to its nearest cluster. Here, $k$ cluster centers will be updated subsequently by minimizing the within-cluster sum of squares (WCSS):
\begin{equation}
    \sum_{i=1}^{k}\sum_{x_i\in C_k}\|x_i-\mu_k\|_2^2,
\end{equation}
where $\mu_k$ is the mean of points locating in the $k$-th cluster $C_k$ and $n_k$ is the number of points in $C_k$. Here, $\| \cdot \|_2$ denotes the $L_2$ distance. 

 The aforementioned algorithm offers an optimal partition of  $k$ clusters. However, it is more important to find the best number of clusters for the given set of SNP variants. Therefore, the Elbow method is employed. A set of WCSS can be calculated in the $k$-means clustering process by varying the number of clusters $k$, and then plot WCSS according to the number of clusters. The optimal number of clusters will be the elbow in this plot. The WCSS measures the variability of the points within each cluster which is influenced by the number of points $N$. Therefore, as the number of total points of $N$ increases, the value of WCSS becomes larger. Additionally, the performance of $k$-means clustering depends on the selection of the specific distance metric.

In this work, we implement $k$-means clustering with the Elbow method for analyzing the optimal number of the subtypes of SARS-CoV-2 SNP variants. The Jaccard distance-based representation is considered as the input features for the $k$-means clustering method. If we have a total of $N$ SNP variants concerning a reference genome in a SARS-CoV-2 sample, the location of the mutation sites for each SNP variant will be saved in the set $S_i, i=1,2,\cdots, N$. The Jaccard distance between two different sets (or samples) $S_i$, $S_j$ is denoted as $d_J(S_i, S_j)$. Therefore, the $N\times N$ Jaccard distance-based representation is 
\begin{equation}
    D_J(i,j) = d_J(S_i, S_j). 
\end{equation}
This representation is used in our $k$-means clustering.

\subsection{Topology-based machine learning prediction of protein-protein binding affinity changes following mutations}

The topology-based network tree (TopNetTree) model was developed by an innovative integration between the topological representation and network tree (NetTree) to predict the binding affinity changes of protein-protein interaction (PPI) following mutation $\Delta\Delta G$ \cite{wang2020topology}. The TopNetTree is applied to predict the binding affinity changes upon mutations that occurred on the RBD of SARS-CoV-2. Algebraic topology \cite{carlsson2009topology} is utilized to simplify the structural complexity of protein-protein complexes and embed vital biological information into topological invariants. NetTree integrates the advantages of convolutional neural networks (CNN) and gradient-boosting trees (GBT), such that CNN is treated as an intermediate model that converts vectorized element- and site-specific persistent homology features into a higher-level abstract feature, and GBT uses the upstream features and other biochemistry features for prediction. The performance test of tenfold cross-validation on the dataset (SKEMPI 2.0 \cite{jankauskaite2019skempi}) carried out using gradient boosted regression trees (GBRTs). The errors with the SKEMPI2.0 dataset are 0.85 in terms of Pearson correlations coefficient ($R_p$)   and 1.11 kcal/mol in terms of the root mean square error (RMSE) \cite{wang2020topology}.

\subsection{Topology-based machine learning prediction of protein folding stability changes following mutation}

In this work, the prediction of protein folding stability changes upon mutation is carried out using a topology-based mutation predictor (TML-MP) ({https://weilab.math.msu.edu/TML/TML-MP/}) which was introduced in literature~\cite{cang2017analysis}. The essential biological information is revealed by persistent homology \cite{carlsson2009topology}. The machine learning features are generated by the element-specific persistent homology and biochemistry techniques. The dataset includes 2648 mutations cases in 131 proteins provided by Dehouck et al~\cite{dehouck2009fast} and is trained by a gradient boosted regression trees (GBRTs). The error with the corresponding dataset is given as Pearson correlations coefficient ($R_p$) of 0.79 and root mean square error (RMSE) of 0.91 kcal/mol from previous work~\cite{cang2017analysis}.

The persistent homology is widely applied in a variety of practical feature generation problems \cite{carlsson2009topology}. It is also successful in the implementation of predictions of protein folding stability changes upon mutation~\cite{cang2017analysis}. The key idea in TML-MP is using the element-specific persistent homology (ESPH) which distinguishes different element types of biomolecules when building persistent homology barcodes. Commonly occurring protein element types include C, N, O, S, and H, where hydrogen and sulfur are excluded according to that hydrogen atoms are often absent from PDB data and sulfur atoms are too few in most proteins to be statistically important. Thus, C, N, and O elements are considered on the ESPH in protein characterization. Features are extracted from the different dimensions of persistent homology barcodes by dividing barcodes into several equally spaced bins which is called binned barcode representation. The auxiliary features, such as geometry, electrostatics, amino acid type composition, and amino acid sequence, are included for machine learning training as well. In TML-MP, gradient boosted regression trees (GBRTs) \cite{friedman2001greedy} are employed to train the dataset according to the size of the training dataset, absence of model overfitting, non-normalization of features, and ability of nonlinear properties ~\cite{cang2017analysis}.

\subsection{Graph network models}
Graph networks can model interactions and their strength between pairs of units in molecules. These approaches are employed to understand mutation-induced structural changes. The biological and chemical properties are measured by comparing descriptors on different networks. In this work, the network consists of a set $S$ of C$_\alpha$ atoms from every residue of protein structure except the target mutation residue such that a C$_\alpha$ atom is included if it is within 16 {\AA } to any atom of the target mutation. The total atom set $T$ is defined as the atoms (C, N, and O) of the target residue and C$_\alpha$ atoms of the network set $S$. Moreover, two vertices are connected in the network if their distance is less than 8 {\AA }. Thus the adjacency matrix $A$ can be defined as well where $A$ is a matrix containing 0 and 1 such that $A(i,j)=0$ if $i$-th and $j$-th atoms are disconnected and $A(i,j)=1$ if $i$-th and $j$-th atoms are connected. Two graph network models employed in this work are described below. 

\textbf{Flexibility-rigidity index (FRI)} FRI  was introduced to study the flexibility of protein molecules \cite{nguyen2016generalized, xia2013multiscale}. The single residue molecular rigidity index measures its influence on the set $S$ which is given as
\begin{equation}
R_\eta^{\alpha} = \sum_{i=1}^{N_S}\sum_{j=1}^{N_T}e^{-\big(\frac{\|\textbf{r}_i-\textbf{r}_j\|}{\eta}\big)^2},
\end{equation}
where $\alpha={\rm w ~or ~\ m}$ stands for the wild type ${\rm w}$ or the mutant type ${\rm m}$, $N_S$ is the number of C$_\alpha$ atoms of the set $S$, and $N_T$ is the number of atoms in total atom set $T$. Here, $\|\textbf{r}_i-\textbf{r}_j\|$ is the distance between atoms at ${\textbf{r}_i}$ and $\textbf{r}_j$. 

\textbf{Average subgraph centrality}

Average subgraph centrality is built on the exponential of the adjacency matrix, $E=e^A$, where $A$ is the aforementioned adjacency matrix. The subgraph centrality is the summation of weighted closed walks of all lengths starting and ending at the same node~\cite{estrada2005subgraph, estrada2020topological}. Thus the average subgraph centrality reveals the average of participating rate of each vertex in all subgraph and the network motif, which is given as
\begin{equation}
\langle C_s^\alpha \rangle = \frac{1}{N_S} \sum_{i=1, i\notin I_T}^{N} E(i,i),
\end{equation}
where $I_T$ is the index set of the mutation residue.

\section{Conclusion}
The prevalence of coronavirus disease 2019 (COVID-19) caused by severe acute respiratory syndrome coronavirus 2 (SARS-CoV-2) in the United States (US) has led to over 4 million infection cases and 145 thousand fatalities as July 22, 2020. Understanding the US SARS-CoV-2 characteristics is of paramount importance to the control of COVID-19 and the reopening of the world's largest economy. 
We genotype the SARS-CoV-2 genome isolates collected since January 2020 and analyze their cluster distributions, co-mutation patterns, and time evolution traits. We show that the US SARS-CoV-2 has evolved into four substrains. We unveil that the top eight US SARS-CoV-2 mutations are split into two essentially disconnected groups. The first group has 5 concurrent mutations that have prevailed over time,  while the other group involving three concurrent mutations is fading out. While five of the top eight US SARS-CoV-2 mutations were initially detected elsewhere  namely, China (2), Singapore (2) and the United Kingdom (1), three of them were homegrown. A wide variety of protein-specific analyses, including sequence alignment, folding stability changes following mutations modeled by persistent homology and machine learning,  molecular flexibility-rigidity index, and averaged subgraph centrality, are employed to investigate the mutation-induced SARS-CoV-2 property changes. We reveal that the first group is associated with stability enhancing or infectivity strengthening mutations, while the second group involves destabilizing mutations.

We demonstrate that overall, genome samples isolated from female patients show more mutations than those isolated from males.  This observation may be due to different strengths of immune response among females and males after the viral infection. We found that one of the top mutations, 27964C$>$T- (S24L on ORF8), has an unusually strong gender dependence. 
    
Finally, we employ algebraic topology and machine learning to predict spike (S) protein and angiotensin-converting enzyme 2 (ACE2) binding affinity changes after mutations.  We show that mutations that strengthen the S protein and ACE2 binding prevail over time. Additionally, three out of four US SARS-CoV-2 substrains have become significantly more infectious. Our results indicate the need for strict SARS-CoV-2 control and containment strategies in the US.

\section*{Supporting Information}
The supporting information is available for 
\begin{enumerate}
    \item[S1:] Supplementary Figures for $k$-means clustering and proteoforms
    \item[S2:] Supplementary Tables for the SNP profiles in the United States and the acknowledgment of GISAID data contributors. 
\end{enumerate}

\section*{Acknowledgment}
This work was supported in part by NIH grant  GM126189, NSF Grants DMS-1721024,  DMS-1761320, and IIS1900473,  Michigan Economic Development Corporation,  George Mason University award PD45722,  Bristol-Myers Squibb, and Pfizer.
The authors thank The IBM TJ Watson Research Center, The COVID-19 High Performance Computing Consortium, NVIDIA, and MSU HPCC for computational assistance.  


\end{document}